\documentclass[12pt]{iopart}

 \usepackage{color}
 \usepackage{epsfig}
 \usepackage{epstopdf}
 \usepackage{iopams}
 \usepackage{cite} 
 
 \expandafter\let\csname equation*\endcsname\relax
 \expandafter\let\csname endequation*\endcsname\relax
 \usepackage{amsmath}
 
\newcommand{\C}[1]{{\cal{#1}}}
\newcommand{\bb}[1]{\textbf{#1}}

\newcommand{\Dz}{\mathcal{D}[\mathbf z]}
\newcommand{\Dzdag}{\mathcal{D}[\mathbf z^\dagger]}
\newcommand{\zb}{\mathbf{z}}
\newcommand{\yb}{\mathbf{y}}
\newcommand{\Dy}{\mathcal{D}[\mathbf y]}
\newcommand{\Dydag}{\mathcal{D}[\mathbf y^\dagger]}
\newcommand{\Pdag}{\mathcal{P}^\dagger[\zb^\dagger]}

\begin{document}

\title[Stochastic thermodynamics based on incomplete information]{Stochastic thermodynamics based on incomplete 
information: Generalized Jarzynski equality with measurement errors with or without feedback }

\author{Christopher W. W\"achtler${}^{1}$, Philipp Strasberg${}^{2}$, and Tobias Brandes}

\address{
{Institut f\"ur Theoretische Physik, Technische Universit\"at Berlin, Hardenbergstr. 36, D-10623 Berlin, Germany}}
\ead{$^1$ christopher.w.waechtler@campus.tu-berlin.de}
\ead{$^2$ phist@physik.tu-berlin.de}
\begin{abstract}
 In the derivation of fluctuation relations, and in stochastic thermodynamics in general, it is tacitly assumed 
 that we can measure the system perfectly, i.e., without measurement errors. We here demonstrate for a driven system
 immersed in a single heat bath, for which the classic Jarzynski equality $\langle e^{-\beta(W-\Delta F)}\rangle = 1$ 
 holds, how to relax this assumption. Based on a general measurement model akin to Bayesian inference we derive a 
 general expression for the fluctuation relation of the \emph{measured} 
 work and we study the case of an overdamped Brownian particle and of a two-level system in particular. We then generalize our results further and incorporate feedback in our description. We show and argue that, if measurement errors are fully taken into account by the agent who controls \emph{and} observes the system, the standard Jarzynski-Sagawa-Ueda relation should be formulated differently. We again 
 explicitly demonstrate this for an overdamped Brownian particle and a two-level system where the fluctuation relation 
 of the measured work differs significantly from the efficacy parameter introduced by Sagawa and Ueda. Instead, the 
 generalized fluctuation relation under feedback control, $\langle e^{-\beta(W-\Delta F)-I}\rangle = 1$, holds only 
 for a superobserver having perfect access to \emph{both} the system and detector degrees of freedom, 
 independently of whether or not the detector yields a noisy measurement record and whether or not we perform feedback.
\end{abstract}


\maketitle

\section{Introduction}

During the last two decades we have seen an enormous progress in the understanding and description of the thermodynamic 
behaviour of small-scale systems, which are strongly fluctuating and arbitrary far from equilibrium. This includes, e.g., 
a consistent thermodynamic description at the single trajectory level and the discovery of so-called fluctuation 
relations which, in a certain sense, promote the status of the second law of thermodynamics from an inequaltiy to an 
equality. A number of excellent review articles and monographs from different perspectives can be found in 
Refs.~\cite{EspositoHarbolaMukamelRMP2009, SekimotoBook2010, CampisiHaenggiTalknerRMP2011, JarzynskiAnnuRevCondMat2011, 
SeifertRPP2012, SchallerBook2014, VandenBroeckEspositoPhysA2015}. 

A tacit assumption underlying this framework, which is usually never discussed in any detail, is that we must be able 
to measure the stochastic trajectory $\bb{z}(t)$ of a system perfectly, i.e., without measurement errors, in order to 
establish the framework of stochastic thermodynamics and to derive fluctuation relations. In practise, we know, however, 
that this is an experimental challenge for very small systems and, to put this thought even further, this might be the 
major obstacle in finding a fully satisfactory generalization of stochastic thermodynamics to quantum systems. 

Extending the framework of stochastic thermodynamics to the case of incomplete or only partially available information 
has only recently attracted interest \cite{RibezziCrivellariRitortPNAS2014, 
AlemanyRibezziCrivellariRitortNJP2015, BechhoeferNJP2015, VissanenEtAlNJP2015, AlonsoLutzRomitoPRL2016, 
GarciaGarciaLahiriLacostePRE2016}. In our context, the results of 
Garc\'{\i}a-Garc\'{\i}a \emph{et al.}~\cite{GarciaGarciaLahiriLacostePRE2016}, who have also derived a modified 
Jarzynski equality for faulty measurements, are of particular importance. Our results are indeed in agreement with their 
theory, though our point of view and derivation differs from them as we will discuss further in the main text below. 

In addition, we also go one step beyond and include feedback based on faulty measurement results in our theory. 
In fact, the state of knowledge of the observer is of crucial importance in control theory and determines how 
``effective'' the feedback control can be applied. However, if the experimentalist is forced to perform feedback based 
on faulty measurement results, it seems logical that she also uses the same (faulty) detector to infer other statistical 
properties of the system. Thus, we argue that, in order to extend stochastic thermodynamics to the case of feedback 
control \emph{with} measurement errors, it is of crucial importance to take this measurement error consistently into 
account also during the time where no feedback is performed but where we still need to measure the system. This has 
indeed crucial consequences as we will examine below. 

\emph{Outline: } The article starts with a derivation of the standard Jarzynski equality (JE) based on a stochastic path 
integral method in order to establish the mathematical tools we will need in the following. Then, the rest of the article 
is divided into two main parts: Sec.~\ref{sec no feedback} treats the case without feedback control and 
Sec.~\ref{sec feedback} the case with feedback control. In both cases we derive a general expression for the measured 
Jarzynski equality (MJE) of the measured work distribution for arbitrary measurement errors [Eqs.~(\ref{eq changed Jarzynski no Feedback eq}) and~(\ref{eq changed Jarzynski feedback eq})]. 
In general, however, these might be extremely difficult to compute. Therefore, we present analytical results 
(underpinned by numerical simulations) for the two paradigmatic cases of an overdamped 
Brownian particle (OBP) in a harmonic potential and a two-level system (TLS). At all times we try to physically motivate our results and shift most lengthy computations to the appendix. Furthermore, we comment on the use of  mutual information in the JE in Sec.~\ref{sec relation to mutual information sec}. Finally, in Sec.~\ref{sec conclusions} we discuss our findings and point out to possible 
future applications.

\section{Derivation of the Jarzynski equality for a driven system in a heat bath}

Consider a system described by a Hamiltonian $H_{\lambda(t)}(z)$. Here, $z$ might denote the position and momentum of a 
particle (i.e., $z = (x,p)$) or the discrete state of a system (such as spin up or down, $z\in\{\uparrow,\downarrow\}$). 
The results derived below are independent of this consideration and we will use the notation of a continuous variable 
$z$ most of the time. Next, suppose the system is in 
contact with a thermal bath at inverse temperature $\beta$ and initially at $t=0$ in equilibrium with it, i.e., 
$p_{t=0}(z) = e^{-\beta H_{\lambda(0)}(z)}/Z_0$ with $Z_0 = \int dz~e^{-\beta H_{\lambda(0)}(z)}$. Then, we change the 
Hamiltonian from $t=0$ to $t=t_f$ as described by an arbitrary but fixed protocol $\lambda(t)$. Consequently, the work performed on the system,
\begin{equation}
\label{eq general work definition eq}
 W = W[\bb z] \equiv \int_0^{t_f} dt~\dot\lambda(t)\frac{\partial H_{\lambda(t)}[\bb z(t)]}{\partial\lambda}, 
\end{equation}
along each trajectory $\bb z(t) = \bb z$ becomes a stochastic quantity whose fluctuations are bounded 
by the following relation, which is also known as Jarzynski's equality (JE)~\cite{JarzynskiPRL1997, JarzynskiPRE1997}, 
\begin{equation}\label{eq Jarznski eq}
 \langle e^{-\beta(W-\Delta F)}\rangle_{\bb z} = 1~.
\end{equation}
Here, $\langle\dots\rangle_{\bb z}$ denotes an average over all possible system trajectories $\bb z$ and 
$\Delta F = -\beta^{-1}(\ln Z_f - \ln Z_0)$ denotes the change in equilibrium free energy. Eq.~(\ref{eq Jarznski eq}) 
can be derived in different ways and we will use stochastic path integrals and the method of time-reversed trajectories 
below. 

In the formalism of stochastic path integrals the average of a trajectory-dependent quantity $X[\mathbf z]$ can be 
expressed as \cite{ChaichianDemichevBook2001}
\begin{equation}
\left<X[\zb] \right>_\zb = \int \mathcal{D}[\mathbf z]~\mathcal P[\mathbf z]~X[\zb] 
\end{equation} 
where $\mathcal{D}[\mathbf z]$ denotes a measure in the space of trajectories $\mathbf{z}$ and 
$\mathcal{P}[\mathbf{z}]$ the probability density (with respect to this measure) of choosing a trajectory $\mathbf{z}$.
We now divide the time interval $[0, t_f]$ into $N$ time steps of duration $\delta t = t_f/N$. A particular trajectory 
$\mathbf{z}$ is then approximated by its coordinates $ z_k\equiv  z(t_k)$ at times $t_k = k \delta t $, $0\leq k \leq N$, such that 
\begin{equation}
\mathbf{z}(t) \to [{z}_0,  z_1, \dots,  z_N] = \mathbf z~.
\end{equation}
Note that the limit $N\to \infty$ by keeping $t_f$ fixed is implied. The work along the trajectory is the discretized version of Eq.~(\ref{eq general work definition eq}),
\begin{equation}
\label{eq general work defintion discretized eq}
W[\zb] = \sum\limits_{k=1}^N \left(H_{\lambda_k}(z_{k-1})-H_{\lambda_{k-1}}(z_{k-1})\right)
\end{equation} 
where $\lambda_k$ denotes the value of the external control parameter at time $t_k$. 
Furthermore, 
\begin{equation}
\int \mathcal{D}[\mathbf z] = \int d z_0\int d z_1 \dots \int d z_N
\end{equation}
where $\int dz$ denotes an integral over a continuous variable (e.g., for an OBP) or a discrete sum 
(e.g., for a TLS). The probability density for a particluar path is given by
\begin{equation}
\mathcal{P}[\mathbf z(t)] = p_{\lambda_0}(z_0)p_{\lambda_1}( z_0\to  z_1)p_{\lambda_2}( z_1\to  z_2)\dots p_{\lambda_N}( z_{N-1}\to  z_N)~.
\end{equation}
Here, $+p_{\lambda_0}(z_0)$ is the initial equilibrium distribution and $p_{\lambda_k}( z_{k-1}\to  z_k)$ denotes the transition 
probability from $ z_{k-1}$ to $ z_k$ in time $\delta t$ where the driving protocol has the value $\lambda_k$. 
This factorization implicitly assumes Markovian system dynamics.

Of particular importance now will be the notion of a time-reversed path, denoted by $\mathbf z^\dagger(t)=[z_N^\ast,z_{N-1}^\ast,...,z_0^\ast]=\zb^\dagger$, 
with time-reversed driving protocol $\lambda^\dagger(t) = \lambda^\ast(t_f-t)$.\footnote{Note that in the presence of a 
magnetic field (or any other odd variable in the Hamiltonian) the sign of the field also changes under time-reversal. }. Here $z_k^\ast$ indicates the time-reversal of $z_k$, e.g., if $z_k=(x_k,p_k)$ for a particle with position $x_k$ and 
momentum $p_k$, then $z_k^\ast = (x_k,-p_k)$. The probability density for such a path is 
\begin{equation}
\mathcal{P}^\dagger[\mathbf z^\dagger] = p_{\lambda^\ast_N}( z_N^\ast)p_{\lambda^\ast_N}(z_N^\ast\to  z_{N-1}^\ast)\dots p_{\lambda^\ast_{1}}( z_1^\ast\to  z_0^\ast)
\end{equation}
As usual in stochastic TD, we assume 
microreversiblity (or local detailed balance)~\cite{CrooksJSP1998, CrooksPRE2000, JarzynskiJSP2000}
\begin{equation}\label{eq microreversibility eq}
p_{\lambda_k^\ast}(z_k^\ast \to  z_{k-1}^\ast) = p_{\lambda_k}( z_{k-1} \to  z_k)e^{\beta \delta q_k(z_{k-1}\to z_k)}
\end{equation}
where  $\delta q_k(z_{k-1}\to z_k) \equiv \delta q_k$ is the heat absorbed by the system during the time interval 
$[t_{k-1},t_k]$. Due to normalization, we can write
\begin{equation}\label{eq Jarzynski derivation eq}
\begin{aligned}
1 &= \int \Dzdag~\mathcal{P}^\dagger[\zb^\dagger]	\\
  &= \int \Dzdag~\frac{p_{\lambda_N^\ast}(z_N^\ast)}{p_{\lambda_0}(z_0)} p_{\lambda_0}(z_0) p_{\lambda_1}(z_0\to z_1) e^{\beta \delta q_1}\dots p_{\lambda_N}(z_{N-1}\to z_N)e^{\beta \delta q_N} \\
	& = \int \Dz~\mathcal{P}[\zb] ~\frac{p_{\lambda_N^\ast}(z_N^\ast)}{p_{\lambda_0}(z_0)} e^{\beta ( \delta q_1 + \dots + \delta q_N)} = \int \Dz~\mathcal{P}[\zb] ~\frac{p_{\lambda_N^\ast}(z_N^\ast)}{p_{\lambda_0}(z_0)} e^{\beta \delta q[\zb]}
\end{aligned}
\end{equation} 
where we used $\Dzdag = \Dz$ and introduced the heat $\delta q[\zb] \equiv \delta q_1+ \dots +\delta q_N$ absorbed along 
the full trajectory $\zb$.
Since the system is initially in equilibrium (in the forward as well as in the backward process), we have
\begin{equation}
\frac{p_{\lambda_N^\ast}(z_N^\ast)}{p_{\lambda_0}(z_0)} = \frac{Z_0}{Z_N}\exp\left[-\beta (H_{\lambda_N^\ast}(z_N^\ast)-H_{\lambda_0}(z_0))\right]
\end{equation}
and furthermore 
$H_{\lambda_N^\ast}(z_N^\ast)-H_{\lambda_0}(z_0)= H_{\lambda_N}(z_N)-H_{\lambda_0}(z_0) \equiv \Delta e(z_0,z_f)$. 
By the first law of thermodynamics the energy difference between initial and final state along the trajectory is 
$\Delta e(z_0,z_f) = q[\zb] + W[\zb]$. Then, from Eq. (\ref{eq Jarzynski derivation eq}) for $N\to \infty$ (keeping $t_f$ fixed) the original JE follows immediately:
\begin{equation}
1 = \int \Dz~\mathcal{P}[\zb]~e^{\beta \Delta F}e^{-\beta W[\zb]} = \left<e^{-\beta (W-\Delta F)}\right>_\zb~.
\end{equation}
To be precise and to emphasize that the statistical average $\langle\dots\rangle_{\bb z}$ is taken over the system 
trajectories we explicitly use a subscript $\bb z$. This will change in the following.

\section{Measured Jarzynski equality without feedback}
\label{sec no feedback}

Suppose now we measure the system coordinate $z$ continuously with measurement outcome $y$, which in general can 
involve measurement errors and suppose the true system dynamics are inaccessible or hidden. 
Then the original JE, evaluated with the accessible measurement data, is in general not equal to unity, but depends on 
the difference of the true and measured work distribution. 

More specifically, we introduce the conditional probability $p_m(y|z)$ to obtain measurement outcome $y$ given a 
particular state $z$ of the system. The probability distribution of measurement outcomes $y$ after a measurement is then 
\begin{equation}
\label{eq definition probability of measured variable eq}
 p'_m(y) = \int dz p_m(y|z)p(z)~.
\end{equation}
Given a particular measurement outcome $y$, the state of the system after the measurement is given by Bayes' rule and 
reads 
\begin{equation}
\label{eq Bayes rule eq}
 p'(z|y) = \frac{p_m(y|z)p(z)}{p'_m(y)}~.
\end{equation}
The case of a perfect measurement, as usually considered in stochastic thermodynamics, is described by 
$p_m(y|z) = \delta_{y,z}$ (where $\delta_{y,z}$ denotes the Kronecker delta for a discrete state space or the Dirac 
distribution for a continuous system). It is then actually redundant to explicitly distinguish between the state of 
the system and the measurement result because $p'_m(y) = p(z=y)$ and $p'(z|y) = \delta_{y,z}$ (the final state is pure 
and coincides with the measurement result). 

\subsection{General case}

\begin{figure}[h]
\begin{center}
\includegraphics[width = 0.49\textwidth]{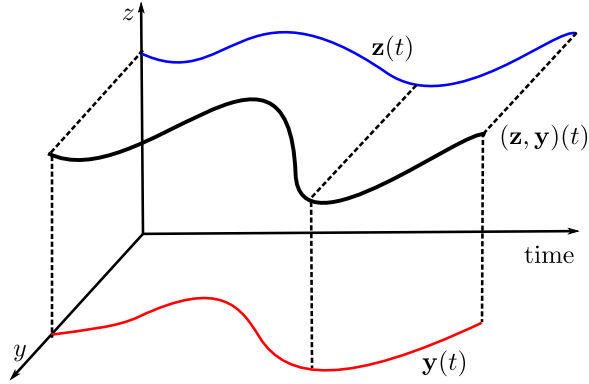}
\end{center}
\caption{Stochastic trajectory $(\zb,\yb)(t)$ (black) in the extended phase space of trajectories. The path $\zb(t)$ of the system (blue) is the projection onto the $z$-subspace and the measured trajectory $\yb(t)$ (red) is the projection onto the subspace of measurement. In general, measured and system trajectories are different.}
\label{fig extension of phase space}
\end{figure}

In order to incorporate the measurements on the system, we expand the phase space to the phase space of measured and true 
trajectories (see Fig.~\ref{fig extension of phase space}). A stochastic path in this extended space is denoted by 
$(\zb, \yb)$ and the probability of choosing such a path is simply denoted by $\mathcal P[\zb,\yb]$. The trajectory 
$\zb$ of the system is the projection of the whole trajectory onto the $z$-subspace and the probability distribution of 
this true stochastic path is given by $\mathcal{P}[\zb]=\int \Dy \mathcal{P}[\zb,\yb]$. Equivalently, the measured 
trajectory $\yb$ lives in the $y$-subspace and its probability distribution is $\mathcal{P}[\yb]=\int \Dz \mathcal{P}[\zb,\yb]$. 
Discretizing the time interval $[0,t_f]$ again into $N$ time steps, the probability density of a path in the space of 
system and measured trajectories will be factorized as 
\begin{equation}
\mathcal{P}[\zb,\yb] = p_{\lambda_0}(z_0,y_0)p_{\lambda_1}(z_0,y_0 \to z_1,y_1)\dots p_{\lambda_{N}}(z_{N-1},y_{N-1}\to z_N, y_N)~.
\end{equation}
Our main assumptions are that the evolution of the system  is independent of the measurement process and that the 
outcome of a measurement $y_k$ only depends on the state of the system $z_k$ at time $t_k$, i.e., we assume 
\begin{equation}
\label{eq factorization of joint probability eq}
p_{\lambda_k}(z_{k-1},y_{k-1}\to z_k,y_k) = p_{\lambda_k}(z_{k-1}\to z_k)p_m(y_k|z_k)~. 
\end{equation}
This can be seen as a Markov assumption for the measurement apparatus, i.e., the previous measurement result $y_{k-1}$ does not 
influence the system evolution and the next measurement result. The conditional probability $p_m(y_k|z_k)$ quantifies 
the uncertainty of the measurement (see Eqs.~(\ref{eq definition probability of measured variable eq}) and~(\ref{eq Bayes rule eq})). 

The measured work $W_m[\yb]$ along a measurement trajectory $\yb$ is defined as in 
Eqs.~(\ref{eq general work definition eq}) and~(\ref{eq general work defintion discretized eq}) by interchanging 
$\zb$ with $\yb$ and is in general \emph{different} from the true work $W = W[\zb]$. Even on average it might be 
that $\langle W_m[\yb]\rangle_\yb \neq \langle W[\zb]\rangle_\zb$. Nevertheless, we assume that the Hamiltonian of the 
system is known to us and unchanged by the measurement; the only mistake is in the measurement outcome $y$ 
(see Ref.~\cite{GarciaGarciaLahiriLacostePRE2016} for the case of different Hamiltonians). 

From an experimental point of view it only makes sense to consider the distribution of measured work and  we may write the average of the exponential of measured work and free energy difference $\Delta F$ as 
\begin{equation}
\label{eq measured JE first step}
\begin{aligned}
\left<e^{-\beta (W_m-\Delta F)}\right>_\yb &= \int \Dy~\mathcal{P}[\yb]~e^{-\beta (W_m[\yb]-\Delta F)} = \int \Dy\Dz~\mathcal{P}[\zb,\yb]~e^{-\beta (W_m[\yb]-\Delta F)} \\
&= \int \Dy\Dz~\mathcal{P}[\zb]\prod\limits_{i=0}^N p_m(y_i|z_i) e^{-\beta (W_m[\yb]-\Delta F)}
\end{aligned}
\end{equation}
where in the last step we used~(\ref{eq factorization of joint probability eq}). 
Again the assumption of microreversibility (see Eq.~(\ref{eq microreversibility eq})) allows us to write Eq.~(\ref{eq measured JE first step}) as
\begin{equation}
\label{eq derivation measured Jarzynski step in between eq}
\left<e^{-\beta (W_m-\Delta F)}\right>_\yb = \int \Dy\Dz~\mathcal{P}^\dagger[\zb^\dagger] \prod\limits_{i=0}^N p_m(y_i|z_i) e^{-\beta \Delta e^\dagger(z_0,z_f)}e^{\beta\delta q^\dagger[\zb^\dagger]}e^{-\beta W_m[\yb]}~. 
\end{equation}
Here, $\Delta e^\dagger(z_0,z_f)$ and $\delta q^\dagger[\zb^\dagger]$ are the energy difference and the exchange of heat with the reservoir along the system's backward trajectory, respectively. The first law also holds for the backwards paths of the system, $\Delta e^\dagger = - \Delta e(z_0,z_f) =  W^\dagger[\zb^\dagger]+\delta q^\dagger[\zb^\dagger]$ and assuming time-reversal symmetry of the measurement, $p_m(y_i|z_i)=p_m(y_i^\ast|z_i^\ast)$, we can further simplify Eq.~(\ref{eq derivation measured Jarzynski step in between eq}) to
\begin{equation}
\begin{aligned}
\left<e^{-\beta (W_m-\Delta F)}\right>_\yb &= \int \Dy\Dz~\mathcal{P}^\dagger[\zb^\dagger] \prod\limits_{i=0}^N p_m(y_i^\ast|z_i^\ast) e^{-\beta W^\dagger[\zb^\dagger]}e^{\beta W_m^\dagger[\yb^\dagger]} \\
&=  \int \Dy\Dz~\mathcal{P}^\dagger[\zb^\dagger,\yb^\dagger]  e^{\beta(W_m^\dagger[\yb^\dagger]-W^\dagger[\zb^\dagger])}
\end{aligned}
\end{equation}
where we have used that the measured work is asymmetric under time reversal, $W^\dagger_m[\yb^\dagger]=-W_m[\yb]$, 
which directly follows from the corresponding property of the true work. Thus, one finally arrives at the 
following expression for the MJE:
\begin{equation}\label{eq changed Jarzynski no Feedback eq}
\left<e^{-\beta (W_m-\Delta F)}\right>_\yb = \left<e^{\beta(W_m^\dagger-W^\dagger)}\right>_{\zb^\dagger,\yb^\dagger}~.
\end{equation}
This expression results from a formal manipulation and is at this point, however, still explicitly dependent 
on the (backward) trajectories $\textbf{z}^\dagger$ of the system and is therefore of limited practical use. Later on we will 
see how to overcome this difficulty for various examples were we use Eq.~(\ref{eq changed Jarzynski no Feedback eq}) as our formal starting point. Note that depending 
on the probability distribution $p(W^\dagger,W_m^\dagger)$ an expansion in terms of the moments of the distribution could be 
also attempted.

As an important limiting case we immediately see that for a perfect measurement, $p_m(y_k|z_k) = \delta_{y_k,z_k}$, 
the measured work coincides with the work of the system, $W_m[\yb] = W[\zb]$, and the right hand side becomes unity 
recovering the original JE (see Eq.~(\ref{eq Jarznski eq})). Moreover, the right hand side of Eq.~(\ref{eq changed Jarzynski no Feedback eq}) may also be equal to one if there is a certain symmetry in the driven system, such that $W_m^\dagger[\yb^\dagger]=W^\dagger[\zb^\dagger]$ (see, e.g., Sec.~\ref{subsec brownian motion no feedback}).

Finally, let us comment on recent work by Garc\'ia-Garc\'ia 
\emph{et al.}~\cite{GarciaGarciaLahiriLacostePRE2016}, who also derive a modified JE including measurement errors and 
which is equivalent to our result, Eq.~(\ref{eq changed Jarzynski no Feedback eq}). However, their point of view as well 
as the derivation differ from the present approach. Garc\'ia-Garc\'ia \emph{et al.} introduce the error $E[\zb,\yb]=W[\zb]-W_m[\yb]$ of system and measured work and derive a fluctuation theorem for the joint distribution of the measured work and this error \cite{GarciaGarciaLahiriLacostePRE2016}:
\begin{equation}
\ln\frac{p'(W_m,E)}{p'^\dagger(-W_m,-E)}= \beta (W_m+E-\Delta F)~.
\end{equation}
From the latter relation, one can immediately derive Eq.~(\ref{eq changed Jarzynski no Feedback eq}). Thus, whereas all measurement errors in Ref.~\cite{GarciaGarciaLahiriLacostePRE2016} are incorporated at the level of the 
final work distribution $p'(W_m,E)$, we \textbf{start with a particular measurement model for the state of the system expressed 
in terms of} $p_m(y_k|z_k)$. This is closer to a microscopic modeling of the situation because any
measurement model for the system $p_m(y_k|z_k)$ will also yield a certain work distribution $p'(W_m,E)$, 
whereas for a given work distribution $p'(W_m,E)$ there might be many different measurement models (and even different 
systems) which yield the same $p'(W_m,E)$. Thus, our findings show a completely different path to derive fluctuation theorems 
in the presence of measurement errors. Whether our approach or the one of Ref. [13] is superior might depend strongly on the 
specific situation and the system under study.

In the following sections we examine two paradigmatic systems for which the right hand side of Eq.~(\ref{eq changed Jarzynski no Feedback eq}) can be evaluated analytically, namely an overdamped Brownian particle (OBP) in a harmonic potential in Sec.~\ref{subsec brownian motion no feedback} and a two-level system (TLS) in Sec.~\ref{subsec two level no feedback}.

\subsection{Overdamped Brownian motion}
\label{subsec brownian motion no feedback}

We consider the overdamped dynamics of a particle in a harmonic potential in one dimension such that the Hamiltonian of the system is only given by the potential energy:
\begin{equation}\label{eq Hamiltonian OBP}
H_{\lambda(t)}(z) = V_{\lambda(t)}(z) = f_{\lambda(t)}(z-\mu_{\lambda(t)})^2~.
\end{equation}
The stiffness  $f_{\lambda(t)}$ as well as the center of the potential $\mu_{\lambda(t)}$ can be altered in time by an 
external driving protocol $\lambda(t)$. To simulate the system dynamics we use the Langevin equation 
\begin{equation}\label{eq Langevin eq}
 \dot z(t) = -\beta D V'_{\lambda(t)}(z) + \sqrt{2D}\xi(t)
\end{equation}
with diffusion constant D, which is related to the friction constant $\gamma$ by the Einstein relation $D= (\beta \gamma)^{-1}$, and Gaussian white noise $\xi(t)$. 

We specify our measurement model by assuming that the measured position of the particle $y_i$ is normally distributed 
around the real position $z_i$ with a standard deviation of $\sigma_m$, 
\begin{equation}\label{eq gaussian measurement eq}
p_m(y_i|z_i) = \frac{1}{\sqrt{2 \pi \sigma_m^2}}\exp\left(-\frac{(z_i-y_i)^2}{2\sigma_m^2}\right)~,
\end{equation}
such that, if $\sigma_m\to 0$, the conditional probability becomes a Dirac distribution and the measured coordinate 
coincides with the true coordinate of the particle. Such a Gaussian measurement model might be a good approximation 
for a noisy measurement without systematic error (i.e., we have $\langle y\rangle = \langle z\rangle$) and simplifies 
a lot analytical calculations. 

Note that the Langevin equation~(\ref{eq Langevin eq}) now merely presents a convenient numerical tool. From the point 
of view of the observer, it has no objective reality unless $\sigma_m=0$. The correct state of knowledge of the 
observer would be indeed described by a stochastic Fokker-Planck equation~\cite{MilburnQSO1996, 
StrasbergSchallerBrandesSFB2016}. 

\paragraph{Continuous driving protocol.} 
Evaluating the general expression, Eq.~(\ref{eq changed Jarzynski no Feedback eq}), for a continuous and piecewise differentiable (c.p.d.) driving protocol $\lambda(t)$ yields 
(see \ref{subsec appendix derivation overdamped dynamics continuous} for the derivation)
\begin{equation}\label{eq changed Jarzynski BD continuous eq}
\left<e^{-\beta (W_m-\Delta F)}\right>_y = e^{-\sigma_m^2 \beta \Delta f}
\end{equation}
where $\Delta f \equiv f_{\lambda(t_f)}-f_{\lambda(0)} $. The right hand side of the above equation equals unity for $\sigma_m = 0$ corresponding to the original JE. Similarly,  if we vary the width of the potential periodically such that $f_{\lambda(0)} = f_{\lambda(t_f)}$, then the original JE is also recovered. However, this attribute is, as far as we know, specific to the model of the overdamped Brownian particle with c.p.d. driving protocol. In general the right hand side will be different from one. Interestingly, shifting the center $\mu_{\lambda(t)}$ of the potential has no effect at all on the MJE. Furthermore, if we define an effective free energy, $\Delta\tilde F \equiv \Delta F + \sigma_m^2 \Delta f$, which may be interpreted as an additonal contribution due to the uncertainty of the measurements, a JE of the form $\left<e^{-\beta (W_m-\Delta\tilde F)}\right>_y = 1$ holds.

\paragraph{Instantaneous change of driving protocol ("quench").} 
We also derive in \ref{subsec appendix derivation overdamped dynamics instantaneous} an analytic expression for the MJE 
for an instantaneous change of the system Hamiltonian at a time $t_m$ (also called a 'quench'). We consider here that 
the position and the width of the parabola is altered at the same time and is constant before and after $t_m$. We find
\begin{equation}\label{eq changed Jarzynski BD instantaneous eq}
\left<e^{-\beta (W_m-\Delta F)}\right>_y = \frac{1}{\sqrt{1+2\beta \sigma^2 \frac{f_{\lambda(0)}}{f_{\lambda(t_f)}}\Delta f}}\exp\left\{\frac{2\beta^2 \sigma_m^2 f_{\lambda(0)}^2\Delta \mu^2}{1+2\beta^2\sigma_m^2\frac{f_{\lambda(0)}}{f_{\lambda(t_f)}}\Delta f}\right\}
\end{equation}
where $\Delta\mu \equiv \mu_{\lambda(t_f)}-\mu_{\lambda(0)}$ is the difference of the center of the parabola before and 
after $t_m$ \footnote{As a side remark note that Eq.~(\ref{eq changed Jarzynski BD continuous eq}) cannot be reproduced from Eq.~(\ref{eq changed Jarzynski BD instantaneous eq}) although a quench can be modeled as
a limit of a series of continuous functions. This has nothing to do with the phenomenon of absolute irreversibility \cite{MurashitaEtAlPRE2014}.
Instead, from our derivation in \ref{subsec appendix derivation overdamped dynamics continuous} it becomes apparent that this procedure would require us to interchange the limit
of the series of continuous functions with an integral, which is only allowed for a \emph{uniformly convergent} series,
but a series of continuous functions converging to a quench (which is not continuous) is not uniformly convergent (but pointwise
instead).}

\paragraph{Numerics.} 
In order to verify our findings we performed Brownian dynamics (BD) simulations and used the 
\textit{weighted ensemble path sampling} algorithm \cite{HuberKimBPJ1996}, which shifts the computational resources 
towards the sampling of rare trajectories, which have the largest impact on the JE. It has been shown that this method 
is statistically exact for a broad class of Markovian stochastic processes \cite{ZhangJasnowZuckermanJCP2010}. 
Please note that we set $\beta \equiv 1$ for all simulations in this paper. 

\begin{figure}[h]
\begin{center}
\includegraphics[width = 0.49\textwidth]{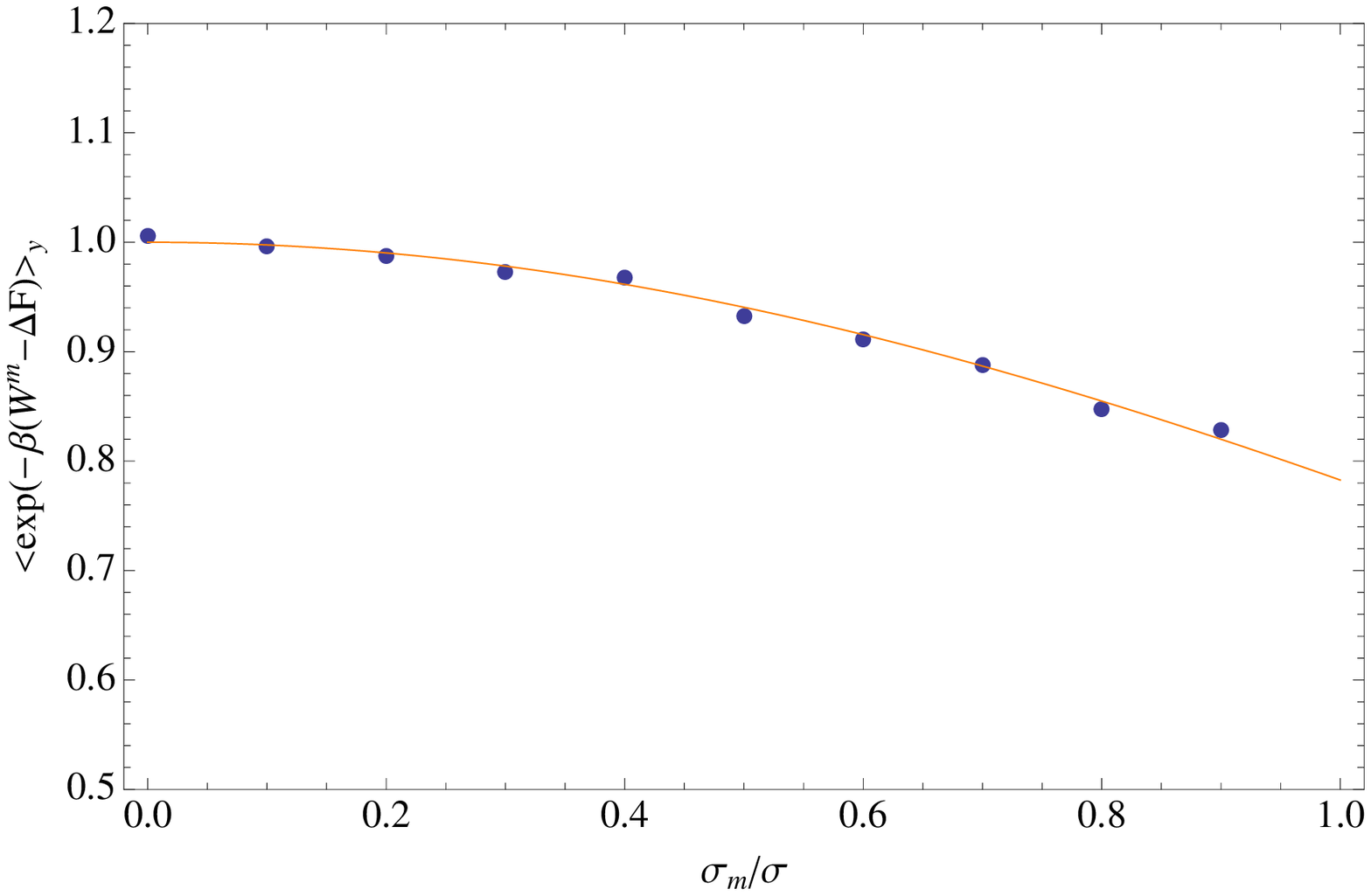}
\includegraphics[width = 0.49\textwidth]{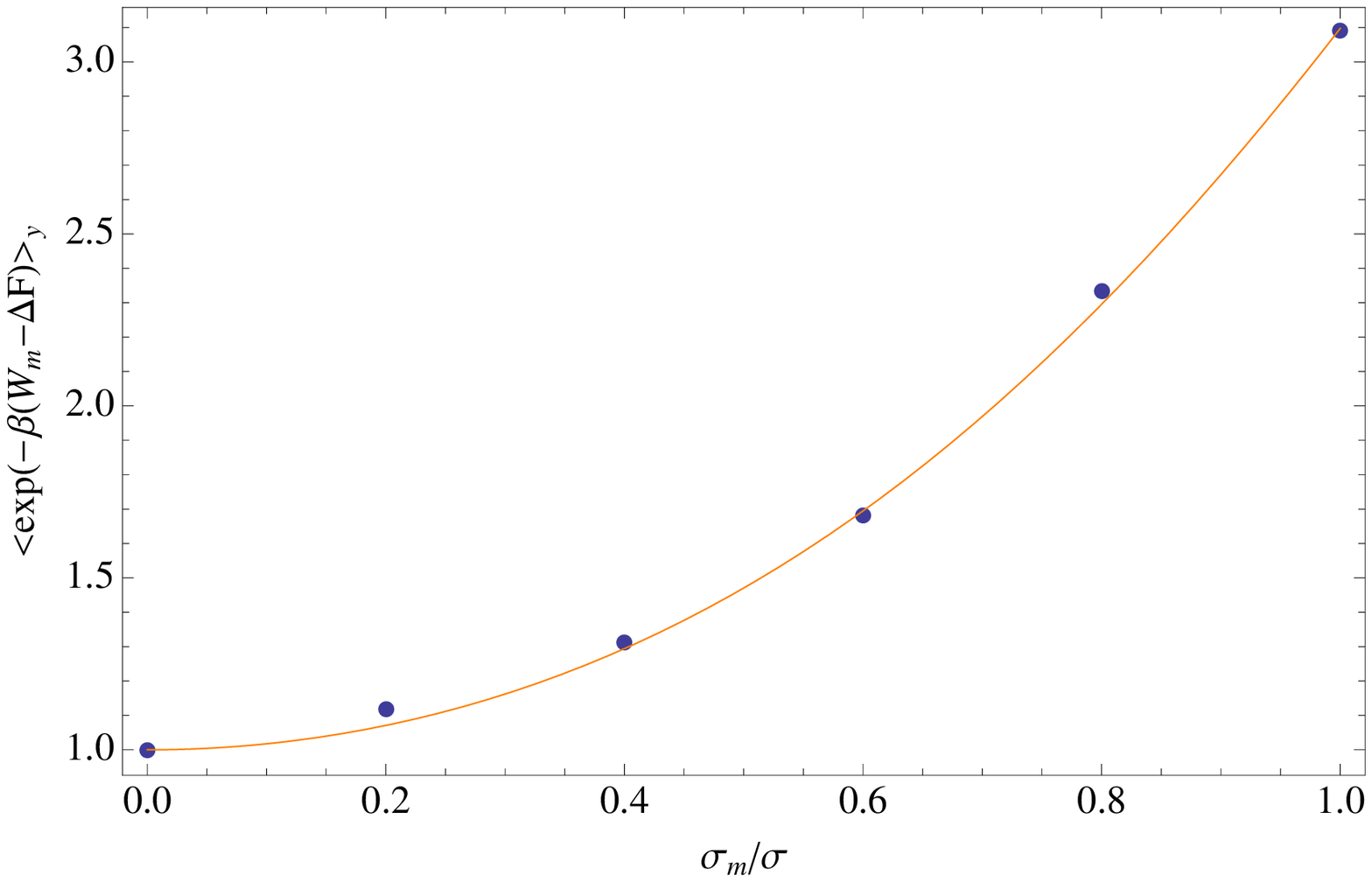}
\end{center}
\caption{Left: Comparison of BD simulation (marks) and the analytic expression 
(Eq.~(\ref{eq changed Jarzynski BD continuous eq}), line) where the system is driven continuously and we choose 
$f_{\lambda(0)}=4.0$, $\alpha = 2.0$, $\mu_{\lambda(0)} =0.0$ and $\alpha' = 2.0$. 
Right: Results of simulation and the analytic expression for a quench of the OBP (see Eq.~(\ref{eq changed Jarzynski BD instantaneous eq})), where $f_{\lambda(0)} = 2.0$, $f_{\lambda(t_f)}=4.0$, $\mu_{\lambda(0)} = 0.0$ and $\mu_{\lambda(t_f)}=1.0$. The quench is performed at time $t_m=t_f/2$. For both driving schemes $t_f=5.0$, $D=2.0$ and we choose $\Delta t = 0.0001$.}
\label{fig comparison BD and anayltics}
\end{figure}

As a simple example we change both parameters of the potential continuously and linearly in time. 
We choose $f_{\lambda(t)} = f_{\lambda(0)} + \alpha t$ and $\mu_{\lambda(t)} = \mu_{\lambda(0)} + \alpha' t$. For this driving 
scheme we find very good agreement of BD simulation and the analytic expression, Eq.~(\ref{eq changed Jarzynski BD continuous eq}), which is presented in Fig.~\ref{fig comparison BD and anayltics} (left).

Furthermore, we compare Eq.~(\ref{eq changed Jarzynski BD instantaneous eq}) with simulation results where intially 
the Hamiltonian of the system is given by $H_0 = f_{\lambda(0)}(z-\mu_{\lambda(0)})^2$ and which is instantaneously changed to 
$H_f = f_{\lambda(t_f)} (z-\mu_{\lambda(t_f)})^2$ at $t_m$. 
In Fig. \ref{fig comparison BD and anayltics} (right) we show the results of the BD simulation (marks) as well as the analytic expression (line) for different values of $\sigma_m$ verifying our findings also for a quench. 

\subsection{Two-level system}
\label{subsec two level no feedback}
Consider a driven system consisting of two energy levels, a ground state with energy $\varepsilon_{\lambda(t)}(g)$ and 
an excited state with energy $\varepsilon_{\lambda(t)}(e)$, coupled to a heat bath with inverse temperature $\beta$. 
The master equation (ME) describing this system is 
\begin{equation}
 \frac{d}{dt}\binom{p_g(t)}{p_e(t)} = \Gamma
 \left(\begin{array}{cc}
        -e^{-\beta\omega_{\lambda(t)}/2}	&	e^{\beta\omega_{\lambda(t)}/2}	\\
        e^{-\beta\omega_{\lambda(t)}/2}		&	-e^{\beta\omega_{\lambda(t)}/2}	\\
       \end{array}\right)\binom{p_g(t)}{p_e(t)}
\end{equation}
Here, we denoted the energy gap of excited and ground state by 
$\omega_{\lambda(t)} \equiv \varepsilon_{\lambda(t)}(e)-\varepsilon_{\lambda(t)}(g)$ and $p_{g/e}(t)$ denotes the 
probability to find the system in the ground/excited state. 

We measure the state of the system continuously with $(1-\eta)$ being the probability of measuring the state of the 
system correctly and consequently $\eta$ of measuring it wrongly, i.e., we set 
$p_m(y_k|z_k) = (1-\eta)\delta_{y_k,z_k}+\eta(1-\delta_{y_k,z_k})$ with $\eta\in[0,1]$. 

\paragraph{Continuous driving protocol.}
The MJE of the TLS, where the external control parameter $\lambda(t)$ is c.p.d., can be well approximated 
by (see \ref{subsec appendix derivation two level no feedback continuous} for the derivation)
\begin{equation}\label{eq changed Jarzynski two level continuous eq}
\left<e^{-\beta(W^m-\Delta F)} \right>_\yb \approx \exp\left(-\eta\beta\int\limits_0^{t_f} dt~\dot\omega_{\lambda^\dagger(t)}\left(p_{ e}^\dagger(t)-p_{ g}^\dagger(t)\right)\right)
\end{equation}
where $p_{ g/e}^\dagger(t)$ denotes the probability 
that the system is in the ground/excited state in the backward process at time $t$, respectively. Furthermore, $\omega_{\lambda^\dagger(t)}$ denotes the energy gap of the TLS. We remark, that for a c.p.d. protocol with nondifferentiable points at $0<t_1<...<t_K<t_f$ we have to split the integral at the respective points as $\int\limits_0^{t_f}dt = \int\limits_0^{t_1}dt+ \int\limits_{t_1}^{t_2}dt+... +\int\limits_{t_K}^{t_f}dt$. 
Moreover, Eq.~(\ref{eq changed Jarzynski two level continuous eq}) is exact up to first order in $\eta$. For higher 
orders (say $\eta^k$) we have to assume that $\C P[z_{i_1},\dots,z_{i_k}] \approx p(z_{i_1})\dots p(z_{i_k})$ which 
seems to be remarkably well justified (see our numerical results below). In fact, though this result strictly holds 
only for slow driving, orders of $\eta^k$ for $k\gg1$ become negligible since $\eta\in[0,1]$, hence, justifying our 
approximation. Furthermore, it is important to note that for the evaluation of the right hand side of 
Eq.~(\ref{eq changed Jarzynski two level continuous eq}) we only need to solve for the \emph{average} evolution of 
the system (as dictated by the master equation); it is not necessary to have access to higher order statistics.

\paragraph{Instantaneous change of driving protocol ("quench").}
For a quench we assume that at $t_m$ with $0 < t_m <t_f$ the energy levels are shifted instantaneously and are held constant before and after. Then, the MJE is given by (see \ref{subsec appendix derivation two level no feedback instantaneous} for the derivation)
\begin{equation}\label{eq changed Jarzynski two level instantaneous eq}
\left<e^{-\beta (W^m-\Delta F)}\right>_\yb = 1-\eta\left[1-p^\dagger_{ g}(t_m)e^{\beta \Delta \omega^\dagger}-p_{ e}^\dagger(t_m)e^{-\beta \Delta \omega^\dagger}\right]
\end{equation}  
where $\Delta \omega^\dagger \equiv \omega_{\lambda^\dagger(t_f)} - \omega_{\lambda^\dagger(0)}$ and $\omega_{\lambda^\dagger(t)}$ is defined as before. Note that both relations for the TLS (Eq.~(\ref{eq changed Jarzynski two level continuous eq}) and (\ref{eq changed Jarzynski two level instantaneous eq})) give the original JE for perfect measurement ($\eta = 0$).

\paragraph{Numerics.}
\begin{figure}[h]
\begin{center}
\includegraphics[width = 0.49\textwidth]{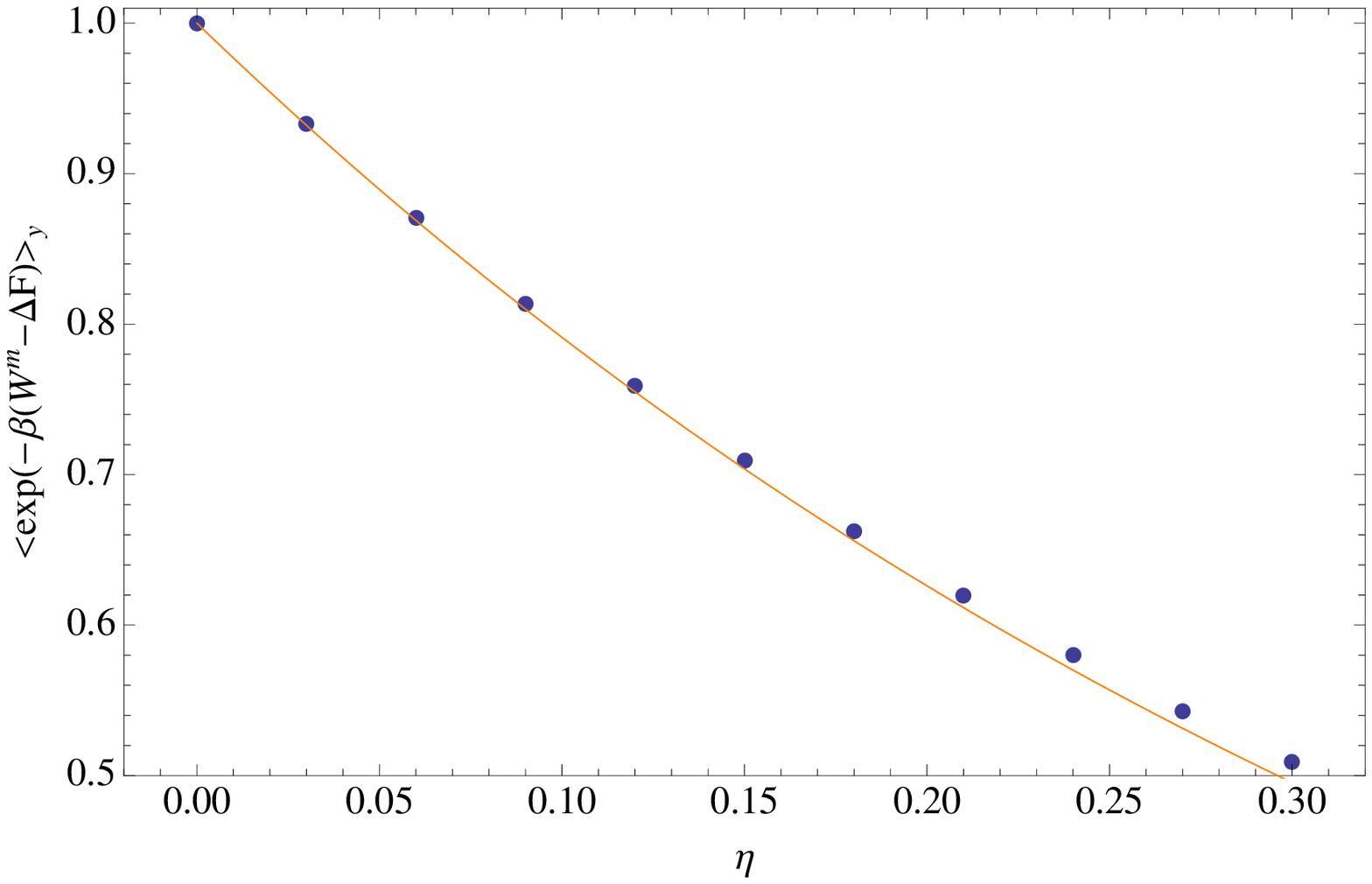}
\includegraphics[width = 0.49\textwidth]{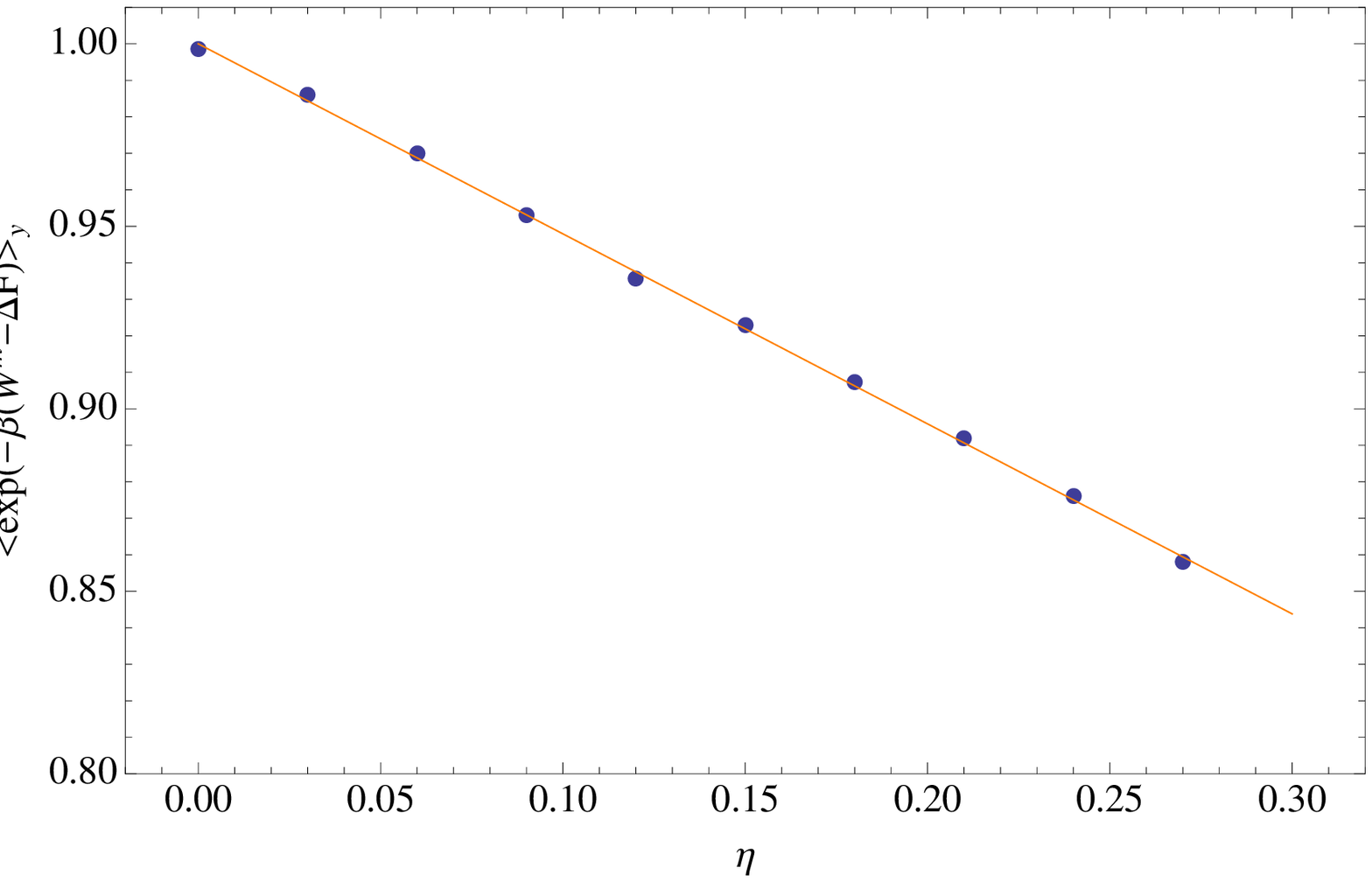}
\caption{Left: Comparison of MC simulation (marks) and numerical integration of the right hand side of Eq.~(\ref{eq changed Jarzynski two level continuous eq}) for a continuously driven TLS, where $\alpha = 1.6\omega_0^2$ and  $\omega_0 \equiv 1$. Right: Simulation results of a quench of the TLS at $t_m=t_f/2$ compared with numerical evaluation of Eq.~(\ref{eq changed Jarzynski two level instantaneous eq}) with $\alpha'=2.0 \omega_0$ and $\omega_0 \equiv 1$. For both driving schemes $\Gamma=10^{-7}/\Delta t$, $\Delta t = 0.001$ and $t_f = 3.0$.}
\label{fig comparison MC and ME two level}
\end{center}
\end{figure}
To test these expression, we performed Monte Carlo (MC) simulations for different values of $\eta \in[0,0.3]$ for two 
driving schemes. First, the driving scheme varies the energy levels continuously and linearly in time, i.e., 
$\omega_{\lambda(t)} = \omega_0+\alpha t$. In Fig.~\ref{fig comparison MC and ME two level} (left) we plotted the left 
hand side of Eq.~(\ref{eq changed Jarzynski two level continuous eq}) from MC simulations (marks) and the right hand 
side from numerical integration of the associated ME of the backward protocol (line). As one can see, 
the approxiamtion of the MJE, Eq.~(\ref{eq changed Jarzynski two level continuous eq}), is in very good agreement with 
the simulation results for small values of $\eta$. Note that a value of $\eta=0.3$ corresponds to a very large error of 
the conditional probability $p_m(y_k|z_k)$ because for a value of $\eta=0.5$ the measurement becomes identical to infering 
the system state by a 
fair coin toss. We also test Eq.~(\ref{eq changed Jarzynski two level instantaneous eq}) where we change the driving protocol instantaneously, i.e., $\omega_{\lambda(t)} = \omega_0+\alpha'\Theta(t-t_m)$. Here, we find perfect agreement of simulation (marks) and numerical integration (line), which is shown in Fig.~\ref{fig comparison MC and ME two level} (right).

\section{Measured Jarzynski equality with feedback}
\label{sec feedback}
Feedback describes the situation in which the state of the system is measured and the evolution of the system is manipulated by applying an external control scheme depending on the measurement outcome. The change of the JE and other fluctuation theorems under feedback has recently attracted a lot of attention, in theory \cite{SagawaUedaPRL2010, PonmuruganPRE2010, HorowitzVaikuntanathanPRE2010, KunduPRE2012, 
LahiriRanaJayannavarJPA2012, AbreuSeifertPRL2012, SagawaUedaPRL2012, SagawaUedaPRE2012, FunoWatanabeUedaPRE2013, 
MunakataRosinbergPRL2014} as well as in experiments \cite{ToyabeEtAlNatPhys2010, KoskiEtAlPRL2014}. 
A prominent and the first example of a generalized JE incorporating feedback by performing a single measurement on a 
stochastic thermodynamic system at a time $t_m$ with measurement outcome $y_m$ is the relation derived by Sagawa and 
Ueda~\cite{SagawaUedaPRL2010}:
\begin{equation}
\label{eq Sagawa and Ueda relation eq}
\left<e^{-\beta (W[\zb|y_m]-\Delta F(y_m))}\right>_{\zb,y_m} = \gamma~.
\end{equation}
The so-called efficacy parameter $\gamma$, which determines 
``how efficiently we use the obtained information with feedback control''~\cite{SagawaUedaPRL2010}, 
depends on the probability $p_{\lambda^\dagger(y_m)}(y_m^\ast)$ of obtaining the time-reversed outcome $y_m^\ast$ 
in the backward process:
\begin{equation}
\label{eq efficacy parameter defintion}
\gamma = \int dy_m~p_{\lambda^\dagger(y_m)} (y_m^\ast)~.
\end{equation}
Note that in the backward process we use the time-reversed driving protocols $\lambda^\dagger(t,y_m)$ according to 
the measurement statistics of $y_m$ obtained in the forward process. Especially, there is no feedback control in the 
backwards process. 

Now, in the derivation of Eq.~(\ref{eq Sagawa and Ueda relation eq}), the particular measurement yielding outcome 
$y_m$ (on which the feedback control is based) is allowed to have measurement errors. However, the left hand 
side of Eq.~(\ref{eq Sagawa and Ueda relation eq}) is evaluated along the \emph{system} trajectories $\zb$, which may be 
inaccessible, especially from an experimental point of view where our knowledge about the situation is solely based on the 
\emph{measurement} trajectories $\yb$. We therefore propose a generalization of the JE under 
feedback control where measurement errors are taken consistently into account. Starting with a general description in 
Sec.~\ref{subsec changed Jarzynski feedback} we look again at the two specific examples of an OBP in a harmonic potential 
including a model of an information ratchet in Sec.~\ref{subsec changed Jarzynski brownian motion feedback} and a feedback controlled 
TLS in Sec.~\ref{subsec changed Jarzynski two level feedback} and verify our analytic results by simulations. Furthermore, in Sec.~\ref{sec relation to mutual information sec} we discuss the relation of the MJE under feedback and the mutual information. 

\subsection{General case}
\label{subsec changed Jarzynski feedback}
Let us suppose we measure our system as we did without feedback control but at one instance in time, denoted $t_m$ with 
$0 < t_m < t_f$, the protocol is changed according to the measurement outcome $y_m$ such that the protocol is fixed 
before $t_m$, i.e., $\lambda = \lambda(t)$ for $t\in[0, t_m]$ and is dependent on $y_m$ after $t_m$, i.e., $\lambda = \lambda(t,y_m)$ for $t\in(t_m, t_f]$. The work applied to the system, which now depends on $y_m$, is given by 
\begin{equation}
\label{eq work definition feedback eq}
W[\zb|y_m] = \int\limits_{0}^{t_m}dt~\dot \lambda(t) \frac{\partial H_{\lambda(t)}[\zb(t)]}{\partial \lambda}+\int\limits_{t_m}^{t_f}dt~\dot\lambda(t,y_m) \frac{\partial H_{\lambda(t,y_m)}[\zb(t)]}{\partial\lambda(y_m)}~.
\end{equation}
The same equation holds also for the measured work $W_m[\yb|y_m]$ by interchanging $\zb$ with $\yb$ (keeping $y_m$). 
The probability of a path in phase space $(\zb,\yb)$ under feedback control is denoted by $\mathcal P_{\lambda(y_m)}[\zb,\yb]$ and we again assume that it factorizes into the probability density of the system trajectory $\mathcal P_{\lambda(y_m)}[\zb]$, which now explicitely depends on $y_m$, and the conditional probabilities $\prod\limits_i p_m(y_i|z_i)$ (see Eq.~(\ref{eq factorization of joint probability eq})). Then, the MJE with feedback control an be expressed as
\begin{equation}
\begin{aligned}
%
\left< e^{-\beta (W_m[\yb|y_m]-\Delta F(y_m))}\right>_\yb &= \int \Dz \Dy~\mathcal{P}_{\lambda(y_m)}[\zb,\yb]~e^{-\beta (W_m[\yb|y_m]-\Delta F(y_m))} \\
%
&= \int \Dz \Dy~\mathcal{P}_{\lambda(y_m)}[\zb]\prod\limits_{i=0}^N p_m(y_i|z_i) e^{-\beta (W_m[\yb|y_m]-\Delta F(y_m))}~.
%
\end{aligned}
\end{equation}
 Note, that the difference in free energy does now also depend on the measurement outcome, i.e., 
 $\Delta F = \Delta F(y_m)$, because the Hamiltonian of the system at time $t_f$ depends on $y_m$. 
 Using again the condition of  microreversiblity (see Eq.~(\ref{eq microreversibility eq})) and assuming time-reversal 
 symmetry of the conditional probabilities, $p_m(y_i|z_i) = p_m(y_i^\ast|z_i^\ast)$, the following equation holds:
 \begin{equation}\label{eq changed Jarzynski feedback step in between eq}
\begin{aligned}
%
&\left< e^{-\beta (W_m[\yb|y_m]-\Delta F(y_m))}\right>_\yb\\
&= \int \Dzdag \Dydag~\mathcal{P}_{\lambda^\dagger(y_m)} [\zb^\dagger] \prod\limits_{i=0}^N p_m(y_i^\ast|z_i^\ast) e^{-\beta (\Delta e^\dagger(y_m^\ast)- \delta q^\dagger(y_m^\ast)+ W_m[\yb|y_m])} \\
%
&= \int \Dzdag \Dydag~\mathcal{P}_{\lambda^\dagger(y_m)}[\zb^\dagger,\yb^\dagger]e^{-W[\zb^\dagger|y_m]}e^{\beta W_m[\yb^\dagger|y_m]}~.
%
\end{aligned}
\end{equation}
From Eq.~(\ref{eq changed Jarzynski feedback step in between eq}) we immediately obtain the MJE in the presence of feedback control:
\begin{equation}
\label{eq changed Jarzynski feedback eq}
\left< e^{-\beta (W_m[\yb|y_m]-\Delta F(y_m))}\right>_\yb= \left<e^{\beta (W_m^\dagger[\yb^\dagger|y_m]-W^\dagger[\zb^\dagger|y_m])}\right>_{\zb^\dagger,\yb^\dagger}~,
\end{equation}
which looks remarkably similar to Eq.~(\ref{eq changed Jarzynski no Feedback eq}). Here, $W_m^\dagger[\yb^\dagger|y_m]$ and $W^\dagger[\zb^\dagger|y_m]$ are the measured and true work, respectively, in the backward process applying the time-reversed protocol $\lambda^\dagger(t,y_m)$ according to the measurement outcome $y_m$ in the forward process. We stress that we do not perform any feedback in the backward process equivalently to \cite{SagawaUedaPRL2010}. Analogously to the efficacy parameter $\gamma$ (see Eqs.~(\ref{eq Sagawa and Ueda relation eq}) and (\ref{eq efficacy parameter defintion})) we call the right hand side of Eq.~(\ref{eq changed Jarzynski feedback eq}) measured efficacy parameter,
\begin{equation}
\label{eq definition measured efficacy eq}
\gamma_m \equiv \left<e^{\beta (W_m^\dagger[\yb^\dagger|y_m]-W^\dagger[\zb^\dagger|y_m])}\right>_{\zb^\dagger,\yb^\dagger}~,
\end{equation}
because the JE is evaluated using the measured trajectories. Note the subtle distinction between Eq.~(\ref{eq Sagawa and Ueda relation eq}) and (\ref{eq changed Jarzynski feedback eq}). Eq.~(\ref{eq Sagawa and Ueda relation eq}) starts with $\left<\exp(-\beta(W-\Delta F))\right>_\zb$ which experimentally requires an error-free detector to evaluate it. We instead start with $\left<\exp(-\beta(W-\Delta F))\right>_\yb$ which can be directly evaluated also with a faulty detector. Our final theoretical result (\ref{eq definition measured efficacy eq}) then depends on $\zb^\dagger$ indeed. However, based on this definition we show below how to overcome this difficulty for various examples. Furthermore, note that a complementary analytical analysis confirming our results has been reported in Ref.~\cite{LahiriArXiv2016} for the example of the Szilard engine.

In the limiting case of perfect measurement, $p_m(y_k|z_k)= \delta_{y_k,z_k}$, Eq.~(\ref{eq changed Jarzynski feedback eq}) simplifies to
\begin{equation}
\label{eq changed Jarzynski feedback perfect measurement step between eq}
\begin{aligned}
%
\left< e^{-\beta W_m[\yb|y_m]-\Delta F(y_m)}\right>_\yb &= \int\Dzdag \Dydag~\mathcal P_{\lambda^\dagger(y_m)}[\zb^\dagger]~\prod\limits_i \delta_{y_i^\ast,z_i^\ast} e^{-\beta(W_m^\dagger[\yb^\dagger|y_m]-W^\dagger[\zb^\dagger|y_m])} \\
%
= \int \Dy\mathcal~P_{\lambda^\dagger(y_m)}[\yb^\dagger] &=\int d y_N^\ast \dots\int d y_0^\ast~p_{\lambda^\ast_N(y_m)}(y^\ast_N)\dots  p_{\lambda^\ast_0}(y_1^\ast\to y_{0}^\ast)~.
%
\end{aligned}
\end{equation}
Due to normalization of conditional probabilites, it holds that the integrals of all $y_k^\ast$ with $k<m$ are equal to unity, hence, 
\begin{equation}
\begin{aligned}
&\int d y_N^\ast \dots\int d y_0^\ast~p_{\lambda_N^\ast(y_m)}(y^\ast_N)\dots  p_{\lambda^\ast_0}(y_1^\ast\to y_{0}^\ast) \\
&=\int d y_N^\ast \dots\int d y_m^\ast~p_{\lambda^\ast_N(y_m)}(y^\ast_N)\dots  p_{\lambda_m^\ast}(y_{m+1}^\ast\to y_{m}^\ast) = \int d y_m^\ast~p_{\lambda^\ast_m(y_m)}(y_m^\ast)\\
&=\int d z^\ast_m~p_{\lambda^\dagger(z_m)}(z_m^\ast)
\end{aligned}
\end{equation}
Only in this case the efficacy $\gamma$ and the measured efficacy $\gamma_m$ are the same as it should be.

However, for a  measurement outcome $y_m$ including 
 errors, $\gamma$ deviates from $\gamma_m$. The interpretation and physical significance of the difference between 
 $\gamma$ and $\gamma_m$ can be explained as follows: consider two observes Alice and Bob. Suppose that Alice measures the state 
 of the system with a faulty detector whereas Bob measures the system with a perfect detector. Furthermore, suppose that only 
 Alice performs the feedback control based on \emph{her} measurement result at time $t_m$. Then, if Alice evaluates the JE of 
 the work done on the system along her measured trajectories, she will observe the result $\gamma_m$. In contrast, Bob -- given 
 the correct system trajectories and knowledge about the feedback action of Alice and her faulty detector -- is able to verify 
 the standard Sagawa-Ueda relation with the efficacy parameter $\gamma$.

\subsection{Overdamped Brownian motion}
\label{subsec changed Jarzynski brownian motion feedback}
As an explicit example, for which we can evaluate the right hand side of Eq.~(\ref{eq changed Jarzynski feedback eq}) 
analytically, we look again at an OBP in a harmonic potential (see Sec.~\ref{subsec brownian motion no feedback}) and 
assume that the center of the potential is intially at $\mu_{\lambda(0)}=0$ and the width is $f_{\lambda(0)}$. Both parameters 
will be changed instantaneously at time $t_m$ if the measured position at that time is $y_m>0$, the position to 
$\mu_{\lambda(t_f)}$ and the stiffness to $f_{\lambda(t_f)}$. Otherwise, for $y_m<0$, the potential remains unchanged. For this 
specific example Eq.~(\ref{eq changed Jarzynski feedback eq}) can be evaluated explicitly and we obtain 
(see~\ref{subsec:AppendixDerivationFeedbackBD})
\begin{equation}
\label{eq brownian motion feedback eq}
\gamma_m = \frac{1}{2}\left(1+\frac{1}{\sqrt{1+ \frac{\kappa}{f_{\lambda(t_f)}}\Delta f}}\text{erfc}\left[-\mu_{t_f}\sqrt{\frac{\beta f_{\lambda(t_f)}(1+\kappa)}{1+ \frac{\kappa}{f_{\lambda(t_f)}}\Delta f}}\right]\exp\left\{\frac{\kappa \beta f_{\lambda(0)} \mu^2_{t_f}}{1+ \frac{\kappa}{f_{\lambda(t_f)}}\Delta f }\right\}\vphantom{\frac{1}{\sqrt{1-2\beta \frac{f_{\lambda(0)}}{f(t_f)}\delta f\sigma_m^2}}}\right)
\end{equation}
where $\kappa \equiv 2 \beta f_{\lambda(0)} \sigma_m^2$.

For the special case of only altering $f_{\lambda(t)}$ and keeping the position of the parabola fixed, i.e.,  $\mu_{\lambda(t_f)}=\mu_{\lambda(0)}$, Eq.~(\ref{eq brownian motion feedback eq}) reduces to
\begin{equation}\label{eq brownian motion feedback breathing parabola eq}
\gamma_m = \frac{1}{2}\left[1+\left(1+2\beta\frac{f_{\lambda(0)}}{f_{\lambda(t_f)}} \Delta f \sigma_m^2\right)^{-1/2}\right]~.
\end{equation}
On the other hand, if the stiffness is held constant, $f_{\lambda(t_f)}=f_{\lambda(0)}=f$, but the parabola is shifted, we find
\begin{equation}\label{eq brownian motion feedback sliding parabola eq}
\gamma_m = \frac{1}{2}\left(1+e^{ 2(f\beta \mu_{t_f}\sigma_m)^2}\text{erfc}\left[-\mu_{t_f}\sqrt{f\beta(1+2f\beta \sigma_m^2)}\right]\right).
\end{equation}
We have varified Eqs.~(\ref{eq brownian motion feedback eq}) - (\ref{eq brownian motion feedback sliding parabola eq}) by perfoming BD simulations for various driving schemes (not shown here) and will discuss the paradigmatic model of an "information ratchet" \cite{SagawaUedaPRL2010} in the next paragraph in more detail also showing numerical results.

\begin{figure}[h]
\begin{center}
\includegraphics[width = 0.49\textwidth]{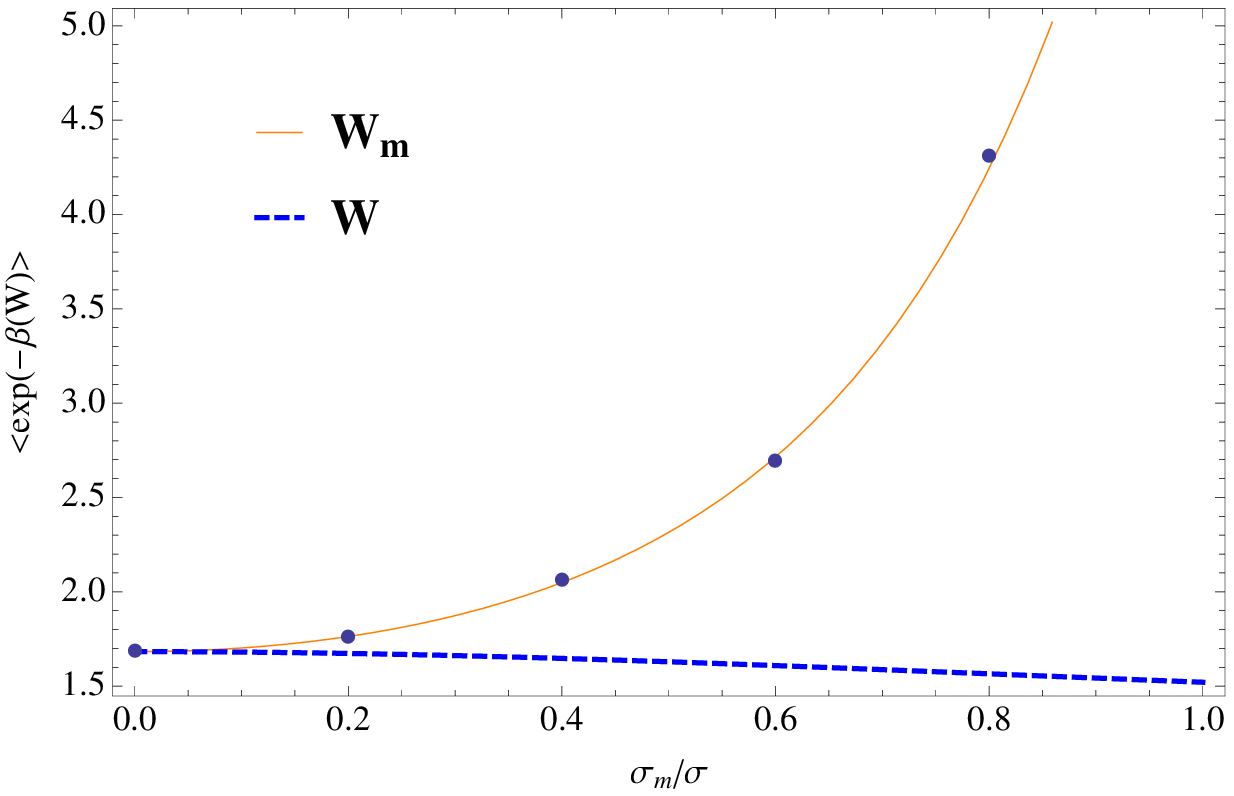}
\includegraphics[width = 0.49\textwidth]{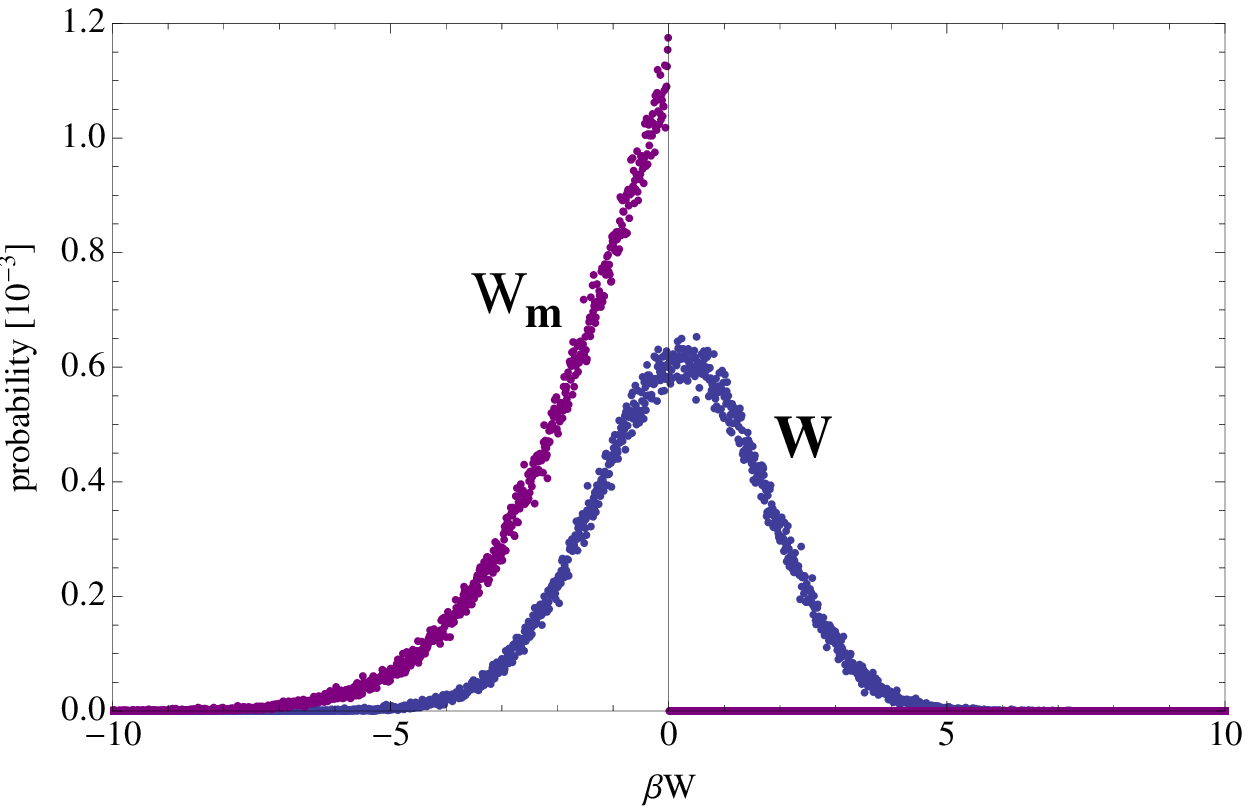}
\end{center}
\caption{Left:  efficacy parameter $\gamma$ (dashed) and measured efficacy $\gamma_m$ (line) as function of the 
measurement error as well as results from BD simulation (marks) with $f=2.0$ and $L=0.5$. Right: Probability distribution of the work extracted 
and performed by the system (blue) and the extracted measured work (purple) for the information ratchet with $\sigma_m/\sigma=1.0$ where $\sigma = (2 \beta f)^{-1/2}$.}
\label{fig Feedback Information Ratchet}
\end{figure}

\paragraph{Information ratchet.} The Brownian particle is initially in thermal equilibrium in the harmonic potential with center $\mu_0$. We then measure 
the position of the particle $y_m$ at time $t_m$ and perform the following feedback scheme: If $y_m\geq \mu_0+L$ with 
$L>0$ being constant, we shift the center of the potential $\mu_{t>t_m} = \mu_0+2L$, if $y_m<\mu_0+L$ we do nothing. 
We then replace $\mu_0\to \mu_0+2L$ and start over again after some transient relaxation time. By repeatingly performing 
this feedback protocol, we can actually move the average position of the particle to the right, ideally without 
performing work. Here, $\Delta F=0$ holds throughout the whole process. Furthermore, one can also extract work from the 
system by this feedback control if the particle is transported against a potential gradient as, e.g., in the 
experiment~\cite{ToyabeEtAlNatPhys2010}. For a single step of the ratchet, where we put $\mu_0=0$ for simplicity, the measured efficacy with feedback control is given by
\begin{equation}\label{eq information ratchet measurements eq}
\gamma_m = \frac{1}{2}\left(\text{erfc}\left[-\frac{L}{\sqrt{\frac{1}{\beta f}+2\sigma_m^2}}\right]
+e^{ 8 f^2 L^2 \beta^2\sigma_m^2}\text{erfc}\left[-L\frac{1+4f \beta \sigma_m^2}{\sqrt{\frac{1}{\beta f}+2\sigma_m^2}}\right]\right)~.
\end{equation}
The derivation follows the same steps as in \ref{subsec:AppendixDerivationFeedbackBD} but the integral of $y_m^\ast$ is splitted at $L$ 
instead of $0$. Eq.~(\ref{eq information ratchet measurements eq}) differs from the efficacy parameter $\gamma$ of the original information ratchet \cite{SagawaUedaPRL2010}, 
\begin{equation}
\gamma = \text{erfc}\left[-\frac{L}{\sqrt{\frac{1}{\beta f}+2\sigma_m^2}}\right]~.
\end{equation}
In Fig. \ref{fig Feedback Information Ratchet} (left) we plot the solutions of the two equations above as function of the variance of the measurement $\sigma_m$.
The two equations coincide for the case of perfect measurement. However, for finite values of $\sigma_m$ the efficacy $\gamma$ of 
the feedback control (dashed line) is lower than for perfect measurement: If the measurement has an error, then the 
potential will be shifted even though the real position of the particle may not be greater than $L$. Then we may 
actually apply work to the system instead of extracting it and the average value of extracted work is lower for noisier 
measurements. 

If we look at the work we measure using the same apparatus as we have used to measure $y_m$ (line), we see that with 
increasing measurement error $\sigma_m$, the measured efficacy $\gamma_m$ also increases in strong contrast to $\gamma$. 
Since the measured work is given in terms of the measured position $y_m$ of the particle, we always apply the ``correct'' 
feedback scheme from the observer's point of view. Thus, we (the observer) always think that we extract work. This can 
also be seen in the distribution of measured (purple) and system (blue) work in 
Fig.~\ref{fig Feedback Information Ratchet} (right), where the probability of measured work is only non-zero for $W_m<0$. 
To support this claim even further, we can calculate the average measured and system work by integration of 
Eq.~(\ref{eq work definition feedback eq}) over $z$ and $y_m$, where the integral is nonzero only if $y_m>L$. The difference 
of them results in 
\begin{equation}
\left<W[\yb|y_m]\right> - \left<W[\zb|y_m]\right> = -4 f L \sigma_m^2\sqrt{\frac{f \beta}{\pi(1+ \kappa)}}\exp\left\{ -\frac{f \beta L^2 }{1+\kappa}\right\} \leq 0
\end{equation}
where $\kappa = 2 \beta f \sigma_m^2$. Thus, on average the measured \emph{extracted} work (note that in our convention 
work is positive if it is done on the system) from the system will be greater than the true extracted work and even 
increases with $\sigma_m$. For a larger value of $\sigma_m$ the probability distribution $p'_m(y_m)$ (see 
Eq.~(\ref{eq definition probability of measured variable eq})) of the measured position $y_m$ is broader (i.e., has a 
larger variance) than $p(z)$, but still has the same mean value as $p(z)$. Then, measurement outcomes with $y_m>L$ are 
more frequent and $\gamma_m$ increases.

\subsection{Two-level system}
\label{subsec changed Jarzynski two level feedback}
Similarly to the derivation of the MJE of the TLS without feedback we find with feedback for a c.p.d. but at this point unspecified driving protocol an approximation for the modification of the original JE (see \ref{subsec appendix two level feedback} for details):
\begin{equation}\label{eq changed Jarzynski feedback two level eq}
\begin{aligned}
\gamma_m &\approx \sum\limits_{z_m\in\{g,e\}} \left[ (1-\eta)p_{\lambda^\dagger(z_m)}(z_m)\exp\left(-\eta\beta \int dt~\dot \omega_{\lambda^\dagger(t,z_m )}\left(p_{ e,\lambda^\dagger(z_m)}(t)-p_{ g,\lambda^\dagger(z_m)}(t)\right)\right)\right.\\
 &\left.-\eta p_{\lambda^\dagger(\bar z_m)}(z_m) \exp\left(-\eta\beta \int dt~\dot \omega_{\lambda^\dagger(t,\bar z_m )}\left(p_{ e,\lambda^\dagger(\bar z_m)}(t)-p_{ g,\lambda^\dagger(\bar z_m)}(t)\right)\right) 	\right]
\end{aligned}
\end{equation}
Here, $p_{z,\lambda^\dagger(y_m)}(t)$ is the probability for the system to be in state $z$ (ground or excited) at time 
$t$ in the backward process with the backward protocol according to the measurement outcome $y_m$ (ground or excited 
state) in the forward process. We again note that we do not apply feedback in the backward process and that 
Eq.~(\ref{eq changed Jarzynski feedback two level eq}) is valid under exactly the same conditions as discussed below 
Eq.~(\ref{eq changed Jarzynski two level continuous eq}). Furthermore, $\omega_{\lambda^\dagger(t,y_m)}$ is the energy gap as defined in Sec.~\ref{subsec two level no feedback} with the time-reversed protocol according to the outcome of the forward process. For a c.p.d. protocol with nondifferentiable points the integral in Eq.~(\ref{eq changed Jarzynski feedback two level eq}) is again split into parts at the respective points. For most driving protocols with feedback we have considered numerically (not shown here) Eq.~(\ref{eq changed Jarzynski feedback two level eq}) is a very good approximation. 

For a driving protocol that is not continuous in time, we find a different expression. Here, we assume as in the case without feedback, that before and after $t_m$ the protocol is constant and that a quench is performed at time $t_m$. We then find for the MJE (see also \ref{subsec appendix two level feedback})
\begin{equation}\label{eq two level system feedback quench eq}
\gamma_m =\sum\limits_{z_m\in \{ g, e\}} \left[(1-\eta) p_{z_m,\lambda^\dagger(z_m)}(t_m) + \eta p_{z_m,\lambda^\dagger(\bar z_m)}(t_m)e^{\beta \Delta \omega_{\lambda^\dagger(\bar z_m)}(z_m)}\right]
\end{equation}
where $p_{z_m,\lambda^\dagger(\bar z_m)}(t_m)$ denotes the probability of the system to be in state $z_m$ at time $t_m$ 
in the backward process with the backward protocol according to the measurement outcome $y_m=\bar z_m$. Here, 
we introduced the complementary state $\bar z_m$ to $z_m$ (i.e., if $z_m=g$ then $\bar z_m = e$ and vice versa). 
Furthermore, $ \Delta \omega_{\lambda^\dagger(\bar z_m)}(z_m) \equiv \omega_{\lambda^\dagger(t_f,\bar z_m)}(z_m) - \omega_{\lambda^\dagger(0,\bar z_m)}(z_m)$ and $\omega_{\lambda^\dagger(t,\bar z_m)}(z_m)=\varepsilon_{\lambda^\dagger(t,\bar z_m)}(z_m) - \varepsilon_{\lambda^\dagger(t,\bar z_m)}(\bar z_m)$. 

We will now discuss an example of a protocol with a quench in detail in the next paragraph.

\paragraph{Conditional swap.} As a specific example, for which we can extract work from a single heat bath by measuring the state of the TLS at time $t_m$, we discuss a feedback operation which we calll a conditional swap: if at time $t_m$ the 
measured state of the TLS $y_m$ is the excited state, we interchange the two energy levels such that we extract work of $\omega = \varepsilon_e-\varepsilon_g$ if the system state $z_m$ is the excited one and perform work of $-\omega$ if $z_m = g$. If $y_m = g$ we do nothing. We compare our findings (see Eq.~(\ref{eq two level system feedback quench eq})) of this conditional swap to the corresponding expression of the efficacy parameter $\gamma$, which is given for this specific example by
\begin{equation}
\gamma = (1-\eta)2 p_{ g,\lambda^\dagger( g)}(t_m)+2\eta p_{ e,\lambda( g)}(t_m) ~.
\end{equation}
Note that in the model of the conditional swap $p_{ g,\lambda^\dagger( g)}(t_m) = p_{ e,\lambda^\dagger( e)}(t_m)$ and $p_{ e,\lambda^\dagger( g)}(t_m)=~p_{ g,\lambda^\dagger( e)}(t_m)$. 

\begin{figure}[h]
\begin{center}
\includegraphics[width = 0.49\textwidth]{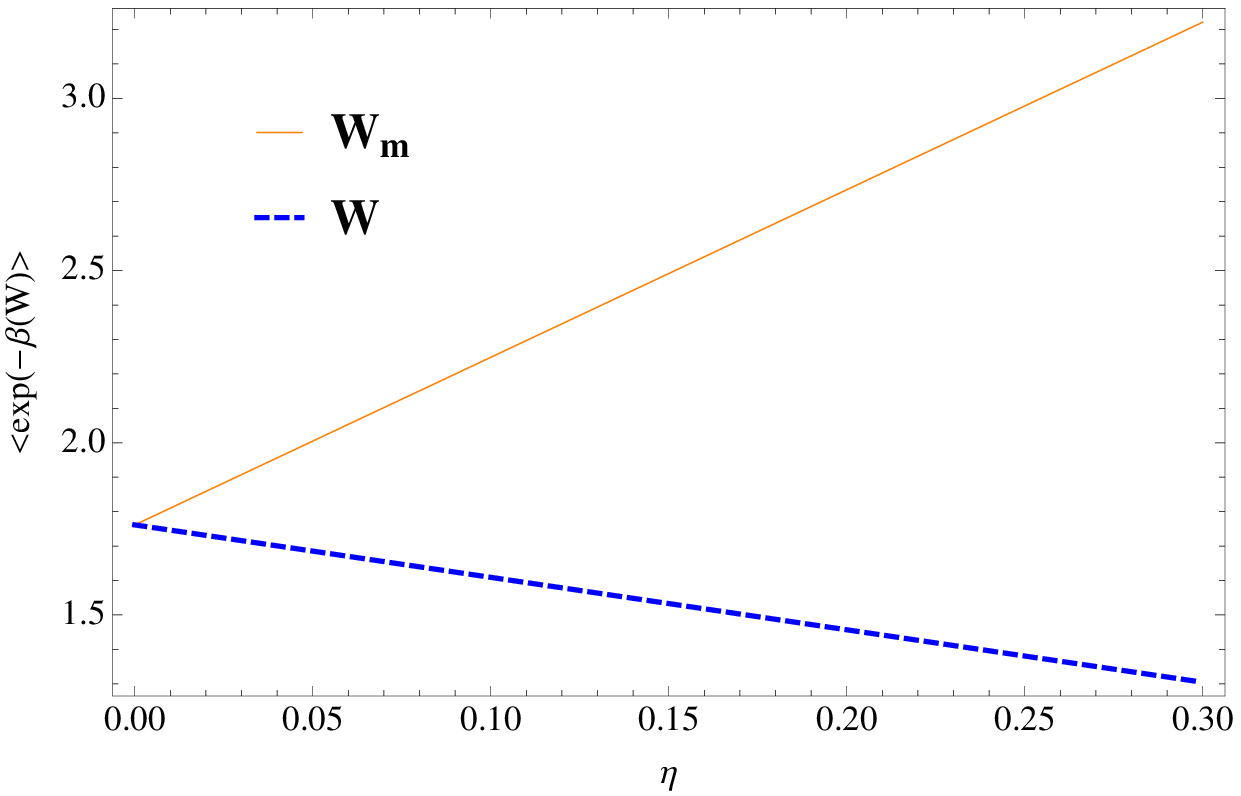}
\includegraphics[width = 0.49\textwidth]{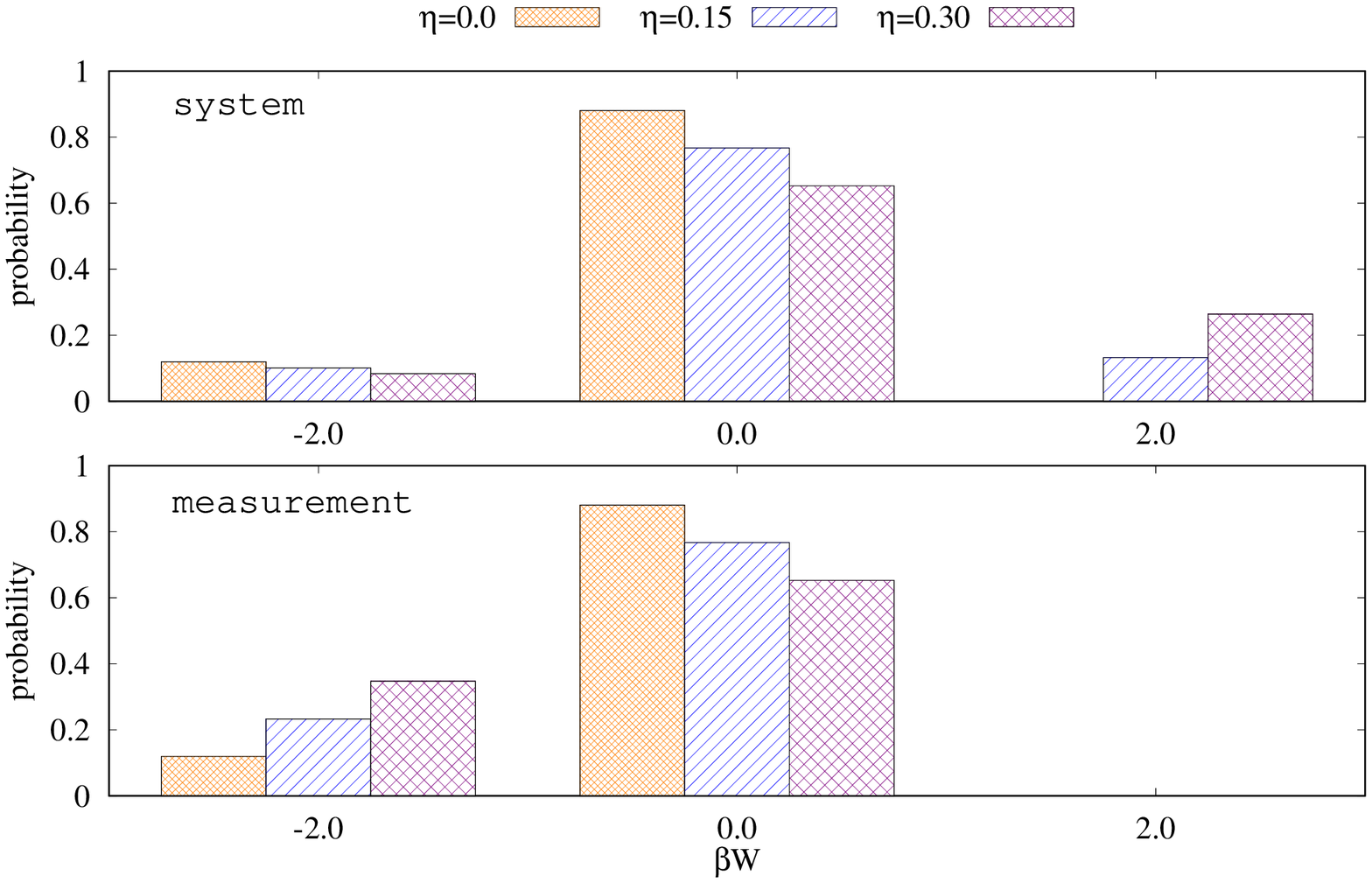}
\end{center}
\caption{Left: Efficacy parameter $\gamma$ of the system (dashed) and measured efficacy $\gamma_m$ (line) as function of the measurement error $\eta$ as well as results from MC simulation (marks) for the conditional swap operation at $t_m = t_f/2$ with $\Gamma=10^{-7}/\Delta t$, $\Delta t = 0.001$ and $t_f = 3.0$. Right: Work distribution of the system (top) and the measured work distribution (bottom) of the conditional swap for different values of $\eta$.}
\label{fig two level system feedback}
\end{figure}

We show the difference of $\gamma$ (dashed) and $\gamma_m$ (line) for different values of $\eta$ in Fig.~\ref{fig two level system feedback} 
(left). As one can see, for a perfect measurement they result in the same value. However if $\eta$ is greater than zero, 
the two differ. The explanation is very similar to the one of the information ratchet discussed in 
Sec.~\ref{subsec changed Jarzynski brownian motion feedback}: if the measurement $y_m$ involves errors, the two states 
are sometimes interchanged even though the system may be in the ground state resulting in work \emph{applied to} the 
system instead of extracting work from the system. If we look at the work distribution of the system (see Fig.~\ref{fig two level system feedback} right top), one can see that for values $\eta > 0$, the extracted work becomes less whereas the probability of applying work to the system increases with measurement error (note that in our convention work is negative if it is done by the system). Then the efficacy parameter is lower than without measurement error. On the other hand, if we look at the measured work (see Fig.~\ref{fig two level system feedback} right bottom), which is calculated from the measured state of the system, we only 
measure positive work extraction from the system by perfoming the conditional swap. Furthermore, the probability of 
measuring the excited state of the system is always larger than the actual probability of the system to be in the excited 
state if $p_e(t_m)<1/2$ (as in our case),
\begin{equation}
p'_{y_m=e}(t_m) = (1-\eta)p_e(t_m)+\eta p_g(t_m) = p_e(t_m)+\eta(1-2 p_e(t_m)) \geq p_e(t_m)~.
\end{equation}
Therefore, the probability of extracting work from the system and therefore $\gamma_m$ increases with larger values of $\eta$. 

\section{Jarzynski equality with mutual information}
\label{sec relation to mutual information sec}

We have seen that the classic JE $\langle e^{-\beta(W-\Delta F)}\rangle = 1$ in general holds only if the system is 
observed perfectly and no feedback is performed. If one of the conditions is violated, we have in general 
$\langle e^{-\beta(W-\Delta F)}\rangle \neq 1$. 
However, in case of feedback at a given time $t_m$ Sagawa and Ueda and others have found that~\cite{SagawaUedaPRL2010, 
PonmuruganPRE2010, HorowitzVaikuntanathanPRE2010, KunduPRE2012, LahiriRanaJayannavarJPA2012, AbreuSeifertPRL2012, 
SagawaUedaPRL2012, SagawaUedaPRE2012, FunoWatanabeUedaPRE2013, MunakataRosinbergPRL2014} 
\begin{equation}\label{eq Jarzynski Sagawa Ueda}
 \langle e^{-\beta(W[\bb z|y_m]-\Delta F(y_m))-I(z_m,y_m)}\rangle_{\bb z,y_m} = 1.
\end{equation}
Thus, by adding the stochastic mutual information $I(z_m,y_m) \equiv \ln\frac{p(y_m,z_m)}{p(y_m)p(z_m)}$ to the exponent 
we can make the right hand side of the ``Jarzynski-Sagawa-Ueda relation'' equal to unity again. This result provides us 
with a nice interpretation because it tells us that the amount of work we can extract from the system is bounded by 
$\langle I(y_m,z_m)\rangle_{z_m,y_m}$, which can be viewed as the amount of correlations established during the 
measurement. 

Unfortunately, in case of measurement errors, validating Eq.~(\ref{eq Jarzynski Sagawa Ueda}) requires to be able to 
observe the system \emph{perfectly} during the time where it is not controlled. But this again raises the question of 
how this might be achieved because this means that the detector of the experimentalist is only faulty previous to the 
feedback step and otherwise correct. 
Eq.~(\ref{eq Jarzynski Sagawa Ueda}) could be therefore viewed as an ``objective'' fluctuation theorem which a second 
``superobserver'' with perfect access to both the system \emph{and} detector degrees of freedom would observe. In 
contrast, the MJE we have considered so far could be called a ``subjective'' fluctuation theorem which is based on the 
knowledge of the observer only. 

In fact, we will now show that by taking the full stochastic mutual information between the system and detector into 
account, defined as 
\begin{equation}
 I[\zb,\yb] = \ln \left(\frac{\mathcal P[\zb,\yb]}{\mathcal P[\zb]P[\yb]}\right),
\end{equation}
yields a fluctuation theorem of the form 
\begin{equation}\label{eq universal FT}
 \langle e^{-\beta(W[\bb z|\bb y]-\Delta F(\bb y))-I(\bb z,\bb y)}\rangle_{\bb z,\bb y} = 1
\end{equation}
which holds without and with measurement errors and without and with feedback, even if the feedback is 
performed continuously, i.e., every time step $\delta t$. However, the latter relation may be invalid for some error-free feedback control processes where absolute irreversiblity is inherent \cite{AshidaEtAlPRE2014}.
We remark that the validity of  Eq.~(\ref{eq universal FT}) without feedback and with measurement errors was already noted in 
Ref.~\cite{GarciaGarciaLahiriLacostePRE2016} and with feedback with or without measurement errors in Refs.~\cite{HorowitzVaikuntanathanPRE2010,LahiriRanaJayannavarJPA2012,SagawaUedaPRE2012}

To prove Eq.~(\ref{eq universal FT}) we note the chain of equalities 
\begin{equation}
 \begin{split}
  \langle e^{-\beta(W[\bb z|\bb y]-\Delta F(\bb y))-I(\bb z,\bb y)}\rangle_{\bb z,\bb y}	&=	\int\Dy \Dz \mathcal{P}[\zb,\yb]e^{-\beta(W_m[\bb z|\bb y]-\Delta F(\bb y))} \frac{\mathcal P[\zb] \mathcal{P}[\yb]}{\mathcal{P}[\zb,\yb]} \\
											&=	\int\Dy \mathcal{P}[\bb y] \int\Dz \mathcal P[\zb] e^{-\beta(W_m[\bb z|\bb y]-\Delta F(\bb y))}	\\
											&=	\int\Dy \mathcal{P}[\bb y] = 1.
 \end{split}
\end{equation}
Here, we used that the JE $\int\Dz \mathcal P[\zb] e^{-\beta(W_m[\bb z|\bb y]-\Delta F(\bb y))} = 1$ holds for every 
fixed measurement record $\bb y$ and (consequently in case of feedback) any control protocol $\lambda(t,\bb y)$. 

Thus, the mutual information seems to be a universal quantity in order to establish fluctuation theorems where not only 
the system but also the detector has to be taken into account, although it does not possess an obvious thermodynamic 
interpretation in case without feedback. Unfortunately, finding some (non-trivial) quantity $G = G[\bb y]$ such that the 
MJE can be corrected, i.e., such that $\langle e^{-\beta(W-\Delta F) - G}\rangle_{\bb y}  = 1$, remains an open problem 
at the moment. 

\section{Conclusions and Outlook}
\label{sec conclusions}

In the present paper, we generalized the original JE expressed in terms of the "true" work done on the system to an 
equation for arbitrary measurement errors based on the measurement record $\yb$. The key ingredient for this was the 
conditional probability distribution $p_m(y|z)$, which quantifies the uncertainty of a measurement outcome $y$ given that 
the system state is $z$ and which defines an abstract measurement model. In fact, by shifting the attention from 
$\zb$ to $\yb$ we only did a first step in generalizing stochastic thermodynamics to the presence of measurement errors 
because much more sophisticated inference schemes could have been considered as well (we actually did not even use 
Eq.~(\ref{eq Bayes rule eq}) in our derivations leaving this interesting problem to future work). 

Then, using the formalism of stochastic path integrals, we derived the MJE (measured JE) without feedback 
(Eq.~(\ref{eq changed Jarzynski no Feedback eq})) and with feedback control 
(Eq.~(\ref{eq changed Jarzynski feedback eq})). These expressions were general (under the assumption of a Markovian 
measurement apparatus), but explicitely involve system trajectory dependent quantities. For two important paradigmatic 
examples we could overcome this difficulty and express the MJE in terms of fixed Hamiltonian parameters or average 
quantities, which can be computed based on a master equation. For an OBP trapped in a harmonic potential the expressions 
derived were exact, whereas for the TLS exact solutions were only found for quenches and very good approximations for 
continuous driving protocols. We also checked our findings with simulation results. In the limiting case of perfect 
measurement the general MJE equations result in the original JE without and with feedback. For the non-ideal 
case we hope that our theory provides a convenient way to explain the always noisy statistics in experiments, 
which have beautifully demonstrated the validity of the JE and other fluctuation theorems within the given statistical 
accuracy so far, see, e.g., Refs.~\cite{HummerSzaboPNAS2001, WangEtAlPRL2002, LiphardtEtAlScience2002, TrepagnierEtAlPNAS2004, 
SchulerEtAlPRL2004, CollinEtAlNature2005, UtsumiEtAlPRB2010, KungEtAlPRX2012, AnEtAlNatPhys2015}. 

Furthermore, in case of feedback control the correct handling of measurement errors is even more important because 
we put the obtained information back into the system to influence its future behaviour. 
Here, we have seen that the measured efficacy $\gamma_m$ may exceed the system efficacy $\gamma$ and, contrary to 
previous intuition, increases with larger measurement errors, which we have calculated explicitely for an information 
ratchet of an OBP and a conditional swap of the TLS. Furthermore, we showed that the "Jarzynski-Sagawa-Ueda relation" by 
incorporating the full stochastic mutual information always holds for a "superobserver" who has access to the measured 
\emph{and} system trajectories, without and with measurement errors and without and with feedback. 

Finally, we would like to mention that a lot of research has already been carried out to understand the stochastic thermodynamics of 
coarse-grained systems, see, e.g., Refs.~\cite{PuglisiEtAlJSM2010, BulnesCuetaraEtAlPRB2011, AltanerVollmerPRL2012, MehlEtAlPRL2012, 
EspositoPRE2012, StrasbergEtAlPRL2013, BulnesCuetaraEtAlPRB2013, LeonardEtAlJCP2013, BaratoHartichSeifertJSP2013, 
ZimmermannSeifertPRE2015, EspositoParrondoPRE2015}. In there, given a set of microstates, a subset of observable states 
is introduced, which defines the coarse-graining and which is sometimes explicitly modeled by a detector or sensor. 
Based on the observability of this subset, the changed laws of (stochastic) thermodynamics are investigated. 
Though one can argue that both approaches pursue the same research goal, it is worthwhile to point out that our 
approach is in principle different. First, the coarse-graining approach still assumes that it is possible to observe 
the particular subsets perfectly, i.e., error-free, and second, it is also implicitly assumed that it is actually possible 
to find these subsets or to physically model a detector, but this might be challenging for some large detectors such as a 
camera. Nevertheless, the question to what extend our approach based on an abstract measurement model $p_m(y|z)$ is equivalent to an explicit detector 
model with underlying coarse-grained system dynamics is, in our point of view, interesting to study in the future.

\section*{Acknowledgments}

Financial support of the DFG through project GRK 1558 is gratefully acknowledged.


\section*{References}

\bibliographystyle{unsrt}
\bibliography{books,open_systems,thermo,info_thermo,general_QM}

\begin{thebibliography}{10}

\bibitem{EspositoHarbolaMukamelRMP2009}
M.~Esposito, U.~Harbola, and S.~Mukamel.
\newblock Nonequilibrium fluctuations, fluctuation theorems, and counting
  statistics in quantum systems.
\newblock {\em Rev. Mod. Phys.}, 81:1665, 2009.

\bibitem{SekimotoBook2010}
K.~Sekimoto.
\newblock {\em Stochastic Energetics}.
\newblock Lect. Notes Phys., Springer, Berlin Heidelberg, 2010.

\bibitem{CampisiHaenggiTalknerRMP2011}
M.~Campisi, P.~H\"anggi, and P.~Talkner.
\newblock Colloquium: Quantum fluctuation relations: Foundations and
  applications.
\newblock {\em Rev. Mod. Phys.}, 83:771, 2011.

\bibitem{JarzynskiAnnuRevCondMat2011}
C.~Jarzynski.
\newblock Equalities and inequalities: irreversibility and the second law of
  thermodynamics at the nanoscale.
\newblock {\em Annu. Rev. Condens. Matter Phys.}, 2:329--351, 2011.

\bibitem{SeifertRPP2012}
U.~Seifert.
\newblock Stochastic thermodynamics, fluctuation theorems and molecular
  machines.
\newblock {\em Rep. Prog. Phys.}, 75:126001, 2012.

\bibitem{SchallerBook2014}
G.~Schaller.
\newblock {\em Open Quantum Systems Far from Equilibrium}.
\newblock Lect. Notes Phys., Springer, Cham, 2014.

\bibitem{VandenBroeckEspositoPhysA2015}
C.~{Van den Broeck} and M.~Esposito.
\newblock Ensemble and trajectory thermodynamics: A brief introduction.
\newblock {\em Physica (Amsterdam)}, 418A:6--16, 2015.

\bibitem{RibezziCrivellariRitortPNAS2014}
M.~Ribezzi-Crivellari and F.~Ritort.
\newblock Free-energy inference from partial work measurements in small
  systems.
\newblock {\em Proc. Natl. Acad. Sci.}, 111:3386--3394, 2014.

\bibitem{AlemanyRibezziCrivellariRitortNJP2015}
A.~Alemany, M.~Ribezzi-Crivellari, and F.~Ritort.
\newblock From free energy measurements to thermodynamic inference in
  nonequilibrium small systems.
\newblock {\em New Journal of Physics}, 17:075009, 2015.

\bibitem{BechhoeferNJP2015}
J.~Bechhoefer.
\newblock Hidden markov models for stochastic thermodynamics.
\newblock {\em New. J. Phys.}, 17:075003, 2015.

\bibitem{VissanenEtAlNJP2015}
K.~L. Viisanen, S.~Suomela, S.~Gasparinetti, O.-P. Saira, J.~Ankerhold, and
  J.~P. Pekola.
\newblock Incomplete measurement of work in a dissipative two level system.
\newblock {\em New. J. Phys.}, 17:055014, 2015.

\bibitem{AlonsoLutzRomitoPRL2016}
J.~J. Alonso, E.~Lutz, and R.~Alessandro.
\newblock Thermodynamics of weakly measured quantum systems.
\newblock {\em Phys. Rev. Lett.}, 116:080403, 2016.

\bibitem{GarciaGarciaLahiriLacostePRE2016}
R.~Garc\'{\i}a-Garc\'{\i}a, L.~Sourabh, and D.~Lacoste.
\newblock Thermodynamic inference based on coarse-grained data or noisy
  measurements.
\newblock {\em Phys. Rev. E}, 93:032103, 2016.

\bibitem{JarzynskiPRL1997}
C.~Jarzynski.
\newblock Nonequilibrium equality for free energy differences.
\newblock {\em Phys. Rev. Lett.}, 78:2690, 1997.

\bibitem{JarzynskiPRE1997}
C.~Jarzynski.
\newblock Equilibrium free-energy differences from nonequilibrium measurements:
  A master-equation approach.
\newblock {\em Phys. Rev. E}, 56:5018--5035, 1997.

\bibitem{ChaichianDemichevBook2001}
M.~Chaichian and A.~Demichev.
\newblock {\em Path Integrals in Physics: Volume II Quantum Field Theory,
  Statistical Physics and other Modern Applications}.
\newblock Institute of Physics, London, 2001.

\bibitem{CrooksJSP1998}
G.~E. Crooks.
\newblock Nonequilibrium measurements of free energy differences for
  microscopically reversible markovian systems.
\newblock {\em J. Stat. Phys.}, 90:1481--1487, 1998.

\bibitem{CrooksPRE2000}
G.~E. Crooks.
\newblock Path-ensemble averages in systems driven far from equilibrium.
\newblock {\em Phys. Rev. E}, 61:2361--2366, 2000.

\bibitem{JarzynskiJSP2000}
C.~Jarzynski.
\newblock Hamiltonian derivation of a detailed fluctuation theorem.
\newblock {\em J. Stat. Phys.}, 98:77--102, 2000.

\bibitem{MilburnQSO1996}
G.~J. Milburn.
\newblock Classical and quantum conditional statistical dynamics.
\newblock {\em Quantum Semiclass. Opt.}, 8:269, 1996.

\bibitem{StrasbergSchallerBrandesSFB2016}
P.~Strasberg, G.~Schaller, and T.~Brandes.
\newblock Controlling the stability of steady states in continuous variable
  quantum systems.
\newblock {\em in \emph{Control of Self-Organizing Nonlinear Systems} (Springer
  International Publishing)}, pages 289--313, 2016.

\bibitem{HuberKimBPJ1996}
G.~A. Huber and S.~Kim.
\newblock Weighted-ensemble brownian dynamics simulations for protein
  association reactions.
\newblock {\em Biophys. J.}, 70:97, 1996.

\bibitem{ZhangJasnowZuckermanJCP2010}
B.~W. Zhang, D.~Jasnow, and D.~M. Zuckerman.
\newblock The "weighted ensemble" path sampling method is statistically exact
  for a broad class of stochastic processes and binning procedures.
\newblock {\em J. Chem. Phys.}, 132:054107, 2010.

\bibitem{SagawaUedaPRL2010}
T.~Sagawa and M.~Ueda.
\newblock Generalized {J}arzynski equality under nonequilibrium feedback
  control.
\newblock {\em Phys. Rev. Lett.}, 104:090602, 2010.

\bibitem{PonmuruganPRE2010}
M.~Ponmurugan.
\newblock Generalized detailed fluctuation theorem under nonequilibrium
  feedback control.
\newblock {\em Phys. Rev. E}, 82:031129, 2010.

\bibitem{HorowitzVaikuntanathanPRE2010}
J.~M. Horowitz and S.~Vaikuntanathan.
\newblock Nonequilibrium detailed fluctuation theorem for repeated discrete
  feedback.
\newblock {\em Phys. Rev. E}, 82:061120, 2010.

\bibitem{KunduPRE2012}
A.~Kundu.
\newblock Nonequilibrium fluctuation theorem for systems under discrete and
  continuous feedback control.
\newblock {\em Phys. Rev. E}, 86:021107, 2012.

\bibitem{LahiriRanaJayannavarJPA2012}
S.~Lahiri, S.~Rana, and A.~M. Jayannavar.
\newblock Fluctuation theorems in the presence of information gain and
  feedback.
\newblock {\em J. Phys. A: Math. Theor.}, 45:065002, 2012.

\bibitem{AbreuSeifertPRL2012}
D.~Abreu and U.~Seifert.
\newblock Thermodynamics of genuine nonequilibrium states under feedback
  control.
\newblock {\em Phys. Rev. Lett.}, 108:030601, 2012.

\bibitem{SagawaUedaPRL2012}
T.~Sagawa and M.~Ueda.
\newblock Fluctuation theorem with information exchange: Role of correlations
  in stochastic thermodynamics.
\newblock {\em Phys. Rev. Lett.}, 109:180602, 2012.

\bibitem{SagawaUedaPRE2012}
T.~Sagawa and M.~Ueda.
\newblock Nonequilibrium thermodynamics of feedback control.
\newblock {\em Phys. Rev. E}, 85:021104, 2012.

\bibitem{FunoWatanabeUedaPRE2013}
K.~Funo, Y.~Watanabe, and M.~Ueda.
\newblock Integral quantum fluctuation theorems under measurement and feedback
  control.
\newblock {\em Phys. Rev. E}, 88:052121, 2013.

\bibitem{MunakataRosinbergPRL2014}
T.~Munakata and M.~L. Rosinberg.
\newblock Entropy production and fluctuation theorems for langevin processes
  under continuous non-{M}arkovian feedback control.
\newblock {\em Phys. Rev. Lett.}, 112:180601, 2014.

\bibitem{ToyabeEtAlNatPhys2010}
S.~Toyabe, T.~Sagawa, M.~Ueda, E.~Muneyuki, and M.~Sano.
\newblock Experimental demonstration of information-to-energy conversion and
  validation of the generalized {J}arzynski equality.
\newblock {\em Nat. Phys.}, 6:988--992, 2010.

\bibitem{KoskiEtAlPRL2014}
J.~V. Koski, V.~F. Maisi, T.~Sagawa, and J.~P. Pekola.
\newblock Experimental observation of the role of mutual information in the
  nonequilibrium dynamics of a {M}axwell demon.
\newblock {\em Phys. Rev. Lett.}, 113:030601, 2014.

\bibitem{HummerSzaboPNAS2001}
G.~Hummer and A.~Szabo.
\newblock Free energy reconstruction from nonequilibrium single-molecule
  pulling experiments.
\newblock {\em Proc. Natl. Acad. Sci.}, 98:3658--3661, 2001.

\bibitem{WangEtAlPRL2002}
G.~M. Wang, E.~M. Sevick, E.~Mittag, D.~J. Searles, and D.~J. Evans.
\newblock Experimental demonstration of violations of the second law of
  thermodynamics for small systems and short time scales.
\newblock {\em Phys. Rev. Lett.}, 89:050601, 2002.

\bibitem{LiphardtEtAlScience2002}
J.~Liphardt, S.~Dumont, S.~B. Smith, I.~Tinoco, and C.~Bustamante.
\newblock Equilibrium information from nonequilibrium measurements in an
  experimental test of {J}arzynski's equality.
\newblock {\em Science}, 296:1832--1835, 2002.

\bibitem{TrepagnierEtAlPNAS2004}
E.~H. Trepagnier, Jarzynski, F.~Ritort, G.~E. Crooks, C.~J. Bustamante, and
  J.~Liphardt.
\newblock Experimental test of {H}atano and {S}asa's nonequilibrium
  steady-state equality.
\newblock {\em Proc. Natl. Acad. Sci.}, 101:15038--15041, 2004.

\bibitem{SchulerEtAlPRL2004}
S.~Schuler, T.~Speck, C.~Tietz, J.~Wrachtrup, and U.~Seifert.
\newblock Experimental test of the fluctuation theorem for a driven two-level
  system with time-dependent rates.
\newblock {\em Phys. Rev. Lett.}, 94:180602, 2005.

\bibitem{CollinEtAlNature2005}
D.~Collin, F.~Ritort, C.~Jarzynski, S.~B. Smith, I.~Tinoco, and C.~Bustamante.
\newblock Verification of the {C}rooks fluctuation theorem and recovery of
  {RNA} folding free energies.
\newblock {\em Nature (London)}, 437:231--234, 2005.

\bibitem{UtsumiEtAlPRB2010}
Y.~Utsumi, D.~S. Golubev, M.~Marthaler, K.~Saito, T.~Fujisawa, and G.~Sch\"on.
\newblock Bidirectional single-electron counting and the fluctuation theorem.
\newblock {\em Phys. Rev. B}, 81:125331, 2010.

\bibitem{KungEtAlPRX2012}
B.~K\"ung, C.~R\"ossler, M.~Beck, M.~Marthaler, D.~S. Golubev, Y.~Utsumi,
  T.~Ihn, and K.~Ensslin.
\newblock Irreversibility on the level of single-electron tunneling.
\newblock {\em Phys. Rev. X}, 2:011001, 2012.

\bibitem{AnEtAlNatPhys2015}
S.~An, J.~N. Zhang, M.~Um, D.~Lv, Y.~Lu, J.~Zhang, Z.-Q. Yin, H.~T. Quan, and
  K.~Kim.
\newblock Experimental test of the quantum {J}arzynski equality with a
  trapped-ion system.
\newblock {\em Nat. Phys.}, 11:193--199, 2015.

\bibitem{PuglisiEtAlJSM2010}
A.~Puglisi, S.~Pigolotti, L.~Rondoni, and A.~Vulpiani.
\newblock Entropy production and coarse graining in markov processes.
\newblock {\em J. Stat. Mech.}, P05015, 2010.

\bibitem{BulnesCuetaraEtAlPRB2011}
G.~Bulnes~Cuetara, M.~Esposito, and P.~Gaspard.
\newblock Fluctuation theorems for capacitively coupled electronic currents.
\newblock {\em Phys. Rev. B}, 84:165114, 2011.

\bibitem{AltanerVollmerPRL2012}
B.~Altaner and J.~Vollmer.
\newblock Fluctuation-preserving coarse graining for biochemical systems.
\newblock {\em Phys. Rev. Lett.}, 108:228101, 2012.

\bibitem{MehlEtAlPRL2012}
J.~Mehl, B.~Lander, C.~Bechinger, V.~Blickle, and U.~Seifert.
\newblock Role of hidden slow degrees of freedom in the fluctuation theorem.
\newblock {\em Phys. Rev. Lett.}, 108:220601, 2012.

\bibitem{EspositoPRE2012}
M.~Esposito.
\newblock Stochastic thermodynamics under coarse graining.
\newblock {\em Phys. Rev. E}, 85:041125, 2012.

\bibitem{StrasbergEtAlPRL2013}
P.~Strasberg, G.~Schaller, T.~Brandes, and M.~Esposito.
\newblock Thermodynamics of a physical model implementing a maxwell demon.
\newblock {\em Phys. Rev. Lett.}, 110:040601, 2013.

\bibitem{BulnesCuetaraEtAlPRB2013}
G.~{Bulnes Cuetara}, M.~Esposito, G.~Schaller, and P.~Gaspard.
\newblock Effective fluctuation theorems for electron transport in a double
  quantum dot coupled to a quantum point contact.
\newblock {\em Phys. Rev. B}, 88:115134, 2013.

\bibitem{LeonardEtAlJCP2013}
T.~Leonard, B.~Lander, U.~Seifert, and T.~Speck.
\newblock Stochastic thermodynamics of fluctuating density fields:
  non-equilibrium free energy differences under coarse-graining.
\newblock {\em J. Chem. Phys.}, 139:204109, 2013.

\bibitem{BaratoHartichSeifertJSP2013}
A.~C. Barato, D.~Hartich, and U.~Seifert.
\newblock Rate of mutual information between coarse-grained non-markovian
  variables.
\newblock {\em J. Stat. Phys.}, 153:460--478, 2013.

\bibitem{ZimmermannSeifertPRE2015}
E.~Zimmermann and U.~Seifert.
\newblock Effective rates from thermodynamically consistent coarse-graining of
  models for molecular motors with probe particles.
\newblock {\em Phys. Rev. E}, 91:022709, 2015.

\bibitem{EspositoParrondoPRE2015}
M.~Esposito and J.~M.~R. Parrondo.
\newblock Stochastic thermodynamics of hidden pumps.
\newblock {\em Phys. Rev. E}, 91:052114, 2015.

\end{thebibliography}

\appendix

\section{Appendix}
\label{sec appendix 1}

\subsection{Derivation of MJE for continuous driving of OBP}
\label{subsec appendix derivation overdamped dynamics continuous}

In this section we derive the analytic expression of the MJE for an OBP in a harmonic potential in one dimension, 
namely Eq.~(\ref{eq changed Jarzynski BD continuous eq}). We assume the external control parameter $\lambda(t)$ to be 
c.p.d. (continuous and piecewise differentiable) throughout this section.
The discretized work along a trajectory $\zb$ given the Hamiltonian in Eq.~(\ref{eq Hamiltonian OBP}) becomes 
\begin{equation}
W[\zb]= \sum\limits_{i} \left( \delta f_{\lambda_i} z_{i-1}^2-2 \delta[ f\mu]_{\lambda_i}z_{i-1}+\delta[ f\mu^2]_{\lambda_i}\right)
\end{equation}
where $\delta f_{\lambda_i} = f_{\lambda_i}-f_{\lambda_{i-1}}$, 
$\delta[ f\mu]_{\lambda_i} = f_{\lambda_i}\mu_{\lambda_i}-f_{\lambda_{i-1}}\mu_{\lambda_{i-1}}$ and 
$\delta[ f\mu^2]_{\lambda_i} = f_{\lambda_i}\mu_{\lambda_i}^2-f_{\lambda_{i-1}}\mu_{\lambda_{i-1}}^2$. 
For the example considered here, it holds that $z_i^\ast = z_i$ and $y_i^\ast = y_i$.

By factorizing $\mathcal{P}[\zb,\yb]$ (see Eq.~(\ref{eq factorization of joint probability eq})) one can express the right hand side of the general Eq.~(\ref{eq changed Jarzynski no Feedback eq}) as 
\begin{equation}
\label{eq appendix starting point BD continuous eq}
\begin{aligned}
%
\left< e^{\beta (W_m^\dagger-W^\dagger)}\right>_{\zb^\dagger,\yb^\dagger} &= \int \Dzdag~\Pdag \int dy _0\dots \int dy _N \prod\limits_{i} \left[p_m(y _i|z _i) \vphantom{\exp\left\{\beta \delta f_{\lambda^\dagger_{i+1}}y^{  2}_i-2\delta[ f\mu]_{\lambda^\dagger_{i+1}}y _i-\delta f_{\lambda^\dagger_{i+1}}z^{  2}_i+2\delta [f\mu]_{\lambda^\dagger_{i+1}}z _i \right\}} \right. \\
%
& \left.\times  \exp\left\{\beta \left(\delta f_{\lambda^\dagger_{i+1}}y^{  2}_i-2\delta[ f\mu]_{\lambda^\dagger_{i+1}}y _i-\delta f_{\lambda^\dagger_{i+1}}z^{  2}_i+2\delta [f\mu]_{\lambda^\dagger_{i+1}}z _i \right)\right\}\right]~.\\
%
\end{aligned}
\end{equation}
Assuming a normal distribution of $p_m(y_i|z_i)$ (see Eq.~(\ref{eq gaussian measurement eq})) we find after integration over all $y_k$:
\begin{equation}
\label{eq appendix BD continuous after integration eq}
\begin{aligned}
%
\left< e^{\beta (W_m^\dagger-W^\dagger)}\right>_{\zb^\dagger,\yb^\dagger} &= \left(\prod\limits_i \frac{1}{\sqrt{1-2\beta \delta f_{\lambda^\dagger_{i+1}}\sigma_m^2}}\right)\times \\
%
& \int \Dzdag~\Pdag\exp\left\{ -\sum\limits_i \frac{2\beta^2\sigma_m^2}{2\beta \delta f_{\lambda^\dagger_{i+1}}\sigma_m^2 -1} \left(\delta[ f\mu]_{\lambda^\dagger_{i+1}}-\delta f_{\lambda^\dagger_{i+1}}z _i\right)^2\right\}~.
%
\end{aligned}
\end{equation}
Note that for the integral over $y_k$ to converge the standard deviation of the measurement must obey
\begin{equation}\label{eq appendix measurement bounded}
\sigma_m^2 < \frac{1}{2\beta \left|\delta f_{\lambda^\dagger_{k+1}}\right|}~.
\end{equation}
This means that in an experimental setup (or also for simulations), in which the width of the potential is varied 
between two measurements by a finite value $\delta f_{\lambda^\dagger_{k+1}}$, the deviation of measured and system coordinate cannot be arbitrarily large.

We first look at the integral of Eq.~(\ref{eq appendix BD continuous after integration eq}): 
in the limit $N\rightarrow\infty$ the time steps $dt = t_f/N$ become infinitesimal and we can write the term in the 
exponential approximately as 
\begin{equation}
 \begin{split}
  & \exp\left\{ -\sum\limits_i \frac{2\beta^2\sigma_m^2}{2\beta \delta f_{\lambda^\dagger_{i+1}}\sigma_m^2 -1} \left(\delta[ f\mu]_{\lambda^\dagger_{i+1}}-\delta f_{\lambda^\dagger_{i+1}}z _i\right)^2\right\}	\\
  & \approx \exp\left\{2\beta^2\sigma_m^2 dt \int_0^{t_f} dt \left([f\mu]'_{\lambda^\dagger(t)}- f'_{\lambda^\dagger(t)}z(t)\right)^2 \right\} \equiv \star
 \end{split}
\end{equation}
where the prime (e.g, $f'$) denotes a derivative with respect to time $t$. Note that the additional $dt$ in front of the 
integral is correct. Furthermore, this step is only exact provided that 
the protocol is differentiable. However, as long as it is continuous and only nondifferentiable at a finite number of 
points $0 < t_1 < \dots < t_K < t_f$ this argument can be easily generalized by splitting the integral at the 
respective places (i.e., $\int_0^{t_1} dt + \int_{t_1}^{t_2} dt + \dots + \int_{t_K}^{t_f} dt$) and by observing that 
due to the continuity $\delta f_{\lambda^\dagger_{i+1}}$ and $\delta[ f\mu]_{\lambda^\dagger_{i+1}}$ remain 
infinitesimal small at all points. Then, by the mean value theorem of integration we know that there exists a 
$\xi\in[0,t_f]$ such that 
\begin{equation}
 \star = \exp\left\{2\beta^2\sigma_m^2 dt \left([f\mu]'_{\lambda^\dagger(\xi)}- f'_{\lambda^\dagger(\xi)}z(\xi)\right)^2 \right\}.
\end{equation}
and hence, this term becomes $1$ for $N\rightarrow\infty$, i.e., $dt\rightarrow0$. 

Therefore, Eq.~(\ref{eq appendix BD continuous after integration eq}) simplifies to
\begin{equation}
\left< e^{\beta (W_m^\dagger-W^\dagger)}\right>_{\zb^\dagger,\yb^\dagger} = \prod\limits_i \frac{1}{\sqrt{1-2\beta \delta f_{\lambda^\dagger_{i+1}}\sigma_m^2}}\approx \prod\limits_i \left(1+\sigma_m^2\beta \delta f_{\lambda^\dagger_{i+1}}\right)~,
\end{equation}
which holds for $N\to \infty$. In the last step, we write the product as an exponential and use an approximation of the logarithm up to first order:
\begin{equation}
\left< e^{\beta (W_m^\dagger-W^\dagger)}\right>_{\zb^\dagger,\yb^\dagger}=  \exp\left(\sum\limits_i\ln\left(1+\sigma_m^2 \beta \delta f_{\lambda^\dagger_{i+1}}\right)\right) \approx \exp\left(\sigma_m^2 \beta (f_{\lambda_0}-f_{\lambda_N})\right)~.
\end{equation}
Taking the limit $N\to\infty$, $f_{\lambda_0} = f_{\lambda(t_0)}$ and $f_{\lambda_N} = f_{\lambda(t_N)}$, we arrive at Eq.~(\ref{eq changed Jarzynski BD continuous eq}).
 
\subsection{Derivation of MJE for instantaneous driving of OBP}
\label{subsec appendix derivation overdamped dynamics instantaneous}

 Here, we derive Eq.~(\ref{eq changed Jarzynski BD instantaneous eq}), where we assume that the stiffness of the harmonic potential as well as the position are instantaneously changed at the same time $t_m$. Since the driving protocol is constant before and after $t_m$, it holds that $\delta f_{\lambda^\dagger_{k+1}} = 0$ as well as $\delta[ f\mu]_{\lambda^\dagger_{k+1}} = 0$  for all $k\neq m$. In this case the right hand side of Eq.~(\ref{eq changed Jarzynski no Feedback eq}) reads after integration over all $y_k$ and $z_k$ with $k\neq m$:
\begin{equation}
\begin{aligned}
%
\left< e^{\beta (W_m^\dagger-W^\dagger)}\right>_{\zb^\dagger,\yb^\dagger} &=\int dz_m ~p^\dagger(z_m ) \int dy_m  \frac{1}{\sqrt{2\pi\sigma_m^2}} \exp\left\{-\frac{(z_m -y_m )^2}{2\sigma_m^2}\right\} \\
%
&   \times\exp\left\{\beta \delta f_{\lambda^\dagger_{m+1}}y^{  2}_m-2\delta[ f\mu]_{\lambda^\dagger_{m+1}}y _m -\delta f_{\lambda^\dagger_{m+1}}z^{  2}_m+2\delta[ f\mu]_{\lambda^\dagger_{m+1}}z _m \right\}.
%
\end{aligned}
\end{equation}
For a quench it holds that $\delta f_{\lambda^\dagger_{m+1}} = f_{\lambda(0)}-f_{\lambda(t_f)} \equiv -\Delta f$ and equivalently $\delta [f\mu]_{\lambda^\dagger_{m+1}}=f_{\lambda(0)}\mu_{\lambda(0)}-f_{\lambda(t_f)}\mu_{\lambda(t_f)} \equiv - \Delta [f\mu]$. Then, the integration over $y_m$ yields
 \begin{equation}
 \begin{aligned}
&\left< e^{\beta (W_m^\dagger-W^\dagger)}\right>_{\zb^\dagger,\yb^\dagger} \\
&= \int dz_m p^\dagger(z_m) \frac{1}{\sqrt{1+2\Delta f\beta \sigma_m^2}} \exp\left\{\frac{2\beta^2\sigma_m^2}{1+2\Delta f\beta \sigma_m^2}\left(-\Delta[ f\mu]+\Delta fz_m\right)^2\right\}~.
 \end{aligned}
 \end{equation}
 For the integral over $y_m$ to converge, it must again hold that $\sigma_m^2< (2\beta \left|\Delta f\right|)^{-1}$. 
 
 We now use, that for the harmonic potential the probability distribution of the position of the OBP in equilibrium (initial system state) is Gaussian distributed with mean $\mu_{\lambda(t_f)}$ and variance $(2 \beta f_{\lambda(t_f)})^{-1/2}$ in the time-reversed protocol. The integration over $z_m$ then finally yields Eq.~(\ref{eq changed Jarzynski BD instantaneous eq}). 
 
 Note that, the integral over $z_m$ only converges if
 \begin{equation}
\sigma_m^2 \leq \frac{1}{2\beta |\Delta f|} \frac{f_{\lambda(t_f)}}{f_{\lambda(0)}}~.
 \end{equation}

\subsection{Derivation of measured Jarzynski equation for a TLS with continuous driving}
\label{subsec appendix derivation two level no feedback continuous}

In this section we derive the analytic expression of the MJE for a  driven TLS, namely 
Eq.~(\ref{eq changed Jarzynski two level continuous eq}). We assume that the protocol $\lambda(t)$ changes continuously
and is piecewise differentiable as in \ref{subsec appendix derivation overdamped dynamics continuous}. For the TLS it 
also holds that $z^\ast = z$ and $y^\ast = y$. The work along a trajectory $\zb$ can be discretized as 
\begin{equation}\label{eq work definition two level eq}
W[\zb] = \sum\limits_{i} \left(\varepsilon_{\lambda_i}(z_{i-1})-\varepsilon_{\lambda_{i-1}}(z_{i-1}) \right) \equiv \sum\limits_i \delta \varepsilon_{\lambda_i}(z_{i-1})~.
\end{equation}
Equivalently, the measured work is given by $W_m[\yb]=\sum\limits_i \delta \varepsilon_{\lambda_i}(y_{i-1})$. 

Then we can evaluate the right hand side of Eq.~(\ref{eq changed Jarzynski no Feedback eq}) analytically as follows:
\begin{equation}
\label{eq appendix start derivation two level continuous eq}
\begin{aligned}
%
&\left< e^{\beta (W_m^\dagger-W^\dagger)}\right>_{\zb^\dagger,\yb^\dagger} \\
%
	& =\sum\limits_{\zb^\dagger} \mathcal{P}^\dagger [\zb^\dagger] \prod_{i} \left(\sum\limits_{y_i} \left[(1-\eta)\delta_{y_i,z_i}+\eta(1-\delta_{y_i,z_i})\right]e^{\beta (\delta \varepsilon_{\lambda^\dagger_{i+1}}(y_i) - \delta \varepsilon_{\lambda^\dagger_{i+1}}(z_i))}\right)\\
%
	&  =\sum\limits_{\zb^\dagger} \mathcal{P}^\dagger [\zb^\dagger] \prod_{i} \left( (1-2\eta)+\eta\left[e^{\beta (\delta \varepsilon_{\lambda^\dagger_{i+1}}( g)-\delta \varepsilon_{\lambda^\dagger_{i+1}}(z_i))}+e^{\beta (\delta \varepsilon_{\lambda^\dagger_{i+1}}(e) - \delta \varepsilon_{\lambda^\dagger_{i+1}}(z_i))}\right] \right)~.
%
\end{aligned}
\end{equation}
Here,  $\sum\limits_{\zb^\dagger}=\sum\limits_{z_N}~\dots~\sum\limits_{z_0}$ denotes all the sums over $z_k$ and 
$\delta \varepsilon_{\lambda^\dagger_k}(z_{k-1})$ is defined as in Eq.~(\ref{eq work definition two level eq}) 
with the time-reversed protocol $\lambda^\dagger(t)$. To further simplify 
Eq.~(\ref{eq appendix start derivation two level continuous eq}) we introduce the complementary state $\bar z_k$ such 
that $\bar z_k \neq z_k$ for all $k$, i.e. if $z_k = e$ then $\bar z_k = g$ and vice versa. Consequently, 
\begin{equation}
\label{eq appendix starting point for derivation instantaneous eq}
\left< e^{\beta (W_m^\dagger-W^\dagger)}\right>_{\zb^\dagger,\yb^\dagger} = \sum\limits_{\zb^\dagger} \mathcal{P}^\dagger [\zb^\dagger] \prod_{i} \left( (1-2 \eta)+\eta \left[1+e^{\beta (\delta \varepsilon_{\lambda^\dagger_{i+1}}(\bar z_i)-\delta \varepsilon_{\lambda^\dagger_{i+1}}( z_i))}\right]\right)~.\\
\end{equation}
For large $N$ we approximate 
\begin{equation}
\label{eq reusing TLS derivation eq 1}
1 +  e^{\beta (\delta \varepsilon_{\lambda^\dagger_{i+1}}(\bar z_i)-\delta \varepsilon_{\lambda^\dagger_{i+1}}( z_i))} \approx 2 +  \beta (\delta \varepsilon_{\lambda^\dagger_{i+1}}(\bar z_i)-\delta \varepsilon_{\lambda^\dagger_{i+1}}( z_i))\equiv 2 + \beta\delta_i~, 
\end{equation}
such that we can write Eq.~(\ref{eq appendix starting point for derivation instantaneous eq}) simply as
\begin{equation}
\begin{aligned}
\left< e^{\beta (W_m^\dagger-W^\dagger)}\right>_{\zb^\dagger,\yb^\dagger} &= \sum\limits_{\zb^\dagger} \mathcal{P}^\dagger [\zb^\dagger] \prod_{i} \left(1+\eta\beta \delta_i\right)~. \\
\end{aligned}
\end{equation}
Writing the product explicitely yields 
\begin{equation}
\left< e^{\beta (W_m^\dagger-W^\dagger)}\right>_{\zb^\dagger,\yb^\dagger}	
 =  \sum\limits_{\zb^\dagger} \mathcal{P}^\dagger [\zb^\dagger] \sum\limits_{n=0}^N\left(\frac{1}{n!}\left(\eta\beta\right)^n \sum\limits_{i_1 \neq ... \neq i_n}\delta_{i_1}\times...\times\delta_{i_n}\right)~.
\end{equation}
We now make the crucial assumption that $\mathcal P^\dagger[z_{k_1},...,z_{k_n}]\approx p^\dagger(z_{k_1})...p^\dagger(z_{k_n})$. 
Then, 
\begin{equation}
\left< e^{\beta (W_m^\dagger-W^\dagger)}\right>_{\zb^\dagger,\yb^\dagger}	
 = \sum\limits_{\zb^\dagger} \sum\limits_{n=0}^N\frac{(\eta\beta)^n}{n!} \sum\limits_{i_1,...,i_n}p^\dagger(z_{i_1})\delta_{i_1}...p^\dagger(z_{i_n})\delta_{i_n} - \mathcal R~. 
\end{equation}
To ensure this equality, we introduced a ``rest'' term $\mathcal R$ of the form
\begin{equation}
 \begin{split}
  \mathcal R	=&~	\frac{(\eta\beta)^2}{2!}\sum\limits_i\sum_{z_i} p^\dagger(z_{i})^2\delta_i^2	\\
		&+	\frac{(\eta\beta)^3}{3!}\sum\limits_{ij}\sum_{z_i,z_j}p^\dagger(z_{i})^2\delta_i^2p^\dagger(z_{j})\delta_j + \frac{(\eta\beta)^3}{3!}\sum\limits_i \sum_{z_i} p^\dagger(z_{i})^3\delta_i^3 + ... ~
 \end{split}
\end{equation}
taking care of the sums where at least two of the indices $i_1,\dots,i_n$ are equal. But then 
all terms of $\mathcal R$ are at least of the order $\mathcal O (\frac{1}{N})$ and therefore vanish for $N\to\infty$. 
Hence, we are left with evaluating 
\begin{equation}
\label{eq appendix right before the end eq}
\left< e^{\beta (W_m^\dagger-W^\dagger)}\right>_{\zb^\dagger,\yb^\dagger} = \sum\limits_{i=0}^N \frac{1}{n!}(\eta\beta)^n \sum\limits_{i_1,..,i_n}\sum\limits_{z_{i_1},...,z_{i_n}} p^\dagger(z_{i_1})\delta_{i_1}...p^\dagger(z_{i_n})\delta_{i_n}~.
\end{equation}
Taking the limit $N\to\infty$, we can write 
\begin{equation}
\lim\limits_{\delta t\to 0}\frac{\delta_k}{\delta t} = \lim\limits_{\delta t\to 0}\frac{\delta\varepsilon_{\lambda^\dagger_{k+1}}(\bar z_k)-\delta \varepsilon_{\lambda_{k+1}^\dagger}(z_k)}{\delta t} = \dot \varepsilon_{\lambda^\dagger(t_{k+1})}(\bar z_k) -\dot \varepsilon_{\lambda^\dagger(t_{k+1})}(z_k)
\end{equation}
where we again assumed that the protocol is differentiable (see the remark below for the case of a c.p.d. 
protocol). 
Evaluating the sums over $z_{i_k}$ and writing the sums over $i_k$ as integrals (by taking $N\to\infty$), 
Eq.~(\ref{eq appendix right before the end eq}) finally reads 
\begin{equation}
\label{eq appendix final step TLS eq}
\left< e^{\beta (W_m^\dagger-W^\dagger)}\right>_{\zb^\dagger,\yb^\dagger} \approx \sum\limits_{n=0}^\infty \frac{1}{n!}(-1)^n \left(\eta \beta \int_0^{t_f} dt~\dot \omega_{\lambda^\dagger(t)}\left(p^\dagger_{e}(t)-p^\dagger_{g}(t)\right)\right)
\end{equation}
where we denote the time derivative of the energy gap of the TLS by 
$\dot \omega_{\lambda^\dagger(t_k)} =\dot \varepsilon_{\lambda^\dagger(t_k)}(e)-\dot\varepsilon_{\lambda^\dagger(t_k)}(g)$ and the probability of the system to be in the ground/exited state at time $t_i$ by $p^\dagger_{g/e}(t_i)$, both in the backward protocol of the driving scheme. Note that Eq.~(\ref{eq appendix final step TLS eq}) is exact up to first order in $\eta$. 

Finally, we remark that for a c.p.d. protocol with nondifferentiable points at 
$0 < t_1 < \dots < t_K < t_f$ the result above readily generalizes and in Eq.~(\ref{eq appendix final step TLS eq}) 
we have to split the integral at the respective points as 
\begin{equation}
\int_0^{t_f} dt = \int_0^{t_1} dt + \int_{t_1}^{t_2} dt + \dots + \int_{t_K}^{t_f} dt~.
\end{equation}

 
\subsection{Derivation of MJE for a TLS for instantaneous driving}
\label{subsec appendix derivation two level no feedback instantaneous}
 
In this section we derive Eq. (\ref{eq changed Jarzynski two level instantaneous eq}), i.e. an expression for the MJE of a TLS, where the energy levels are changed instantaneously at one moment in time $t_m$ with $0<t_m<t_f$ and are constant before and after. Since the energy levels are constant before and after $t_m$, it follows that $\delta\varepsilon_{\lambda^\dagger_{i+1}}(z_i)=0$ for all $i\neq m$ and also  $\delta\varepsilon_{\lambda^\dagger_{i+1}}(\bar z_i) = 0$ for all $i\neq m$. Then the right hand side of Eq.~(\ref{eq changed Jarzynski no Feedback eq}) simplifies to
\begin{equation}
\label{eq appendix derivation TLS no feedback eq}
\begin{aligned}
%
\left< e^{\beta (W_m^\dagger-W^\dagger)}\right>_{\zb^\dagger,\yb^\dagger} &= \sum\limits_{\zb^\dagger}\mathcal P^\dagger[\zb^\dagger]~\left[1-\eta\left(1-e^{\beta(\delta \varepsilon_{\lambda^\dagger_{m+1}}(z_m)-\delta \varepsilon_{\lambda^\dagger_{m+1}}(\bar z_m))}\right)\right] \\ 
%
& = \sum\limits_{z_m\in\{g,e\}}p^\dagger_{z_m}(t_m)\left[1-\eta\left(1-e^{\beta(\delta \varepsilon_{\lambda^\dagger_{m+1}}(z_m)-\delta \varepsilon_{\lambda^\dagger_{m+1}}(\bar z_m))}\right)\right]~.
\end{aligned}
\end{equation}
Summing over $z_m$, Eq.~(\ref{eq appendix derivation TLS no feedback eq}) can be written as
\begin{equation}
\left< e^{\beta (W_m^\dagger-W^\dagger)}\right>_{\zb^\dagger,\yb^\dagger} =1-\eta\left(1-p^\dagger_g(t_m)e^{\beta \Delta \omega^\dagger}-p^\dagger_e(t_m)e^{-\beta \Delta \omega^\dagger}\right)
\end{equation}
 where $\Delta \omega^\dagger = \omega_{\lambda^\dagger(t_f)} - \omega_{\lambda^\dagger(0)}$. Note that this equation is exact for $N\to\infty$ ($\delta t\to 0$).

\subsection{Derivation of the Brownian particle under feedback}\label{subsec:AppendixDerivationFeedbackBD}
For the derivation of Eq.~(\ref{eq brownian motion feedback eq}), the MJE under feedback, we assume that $\mu_{\lambda(0)}=0$ initially and changes instantaneously at $t_m$ to $\mu_{\lambda(t_f)}$ if $y_m>0$. Similarly, the width $f_{\lambda(t)}$ changes from $f_{\lambda(0)}$ to $f_{\lambda(t_f)}$ instantaneously if $y_m>0$. Since the form and the position of the potential is fixed before and after applying the feedback, it holds $\delta H_{\lambda_{k+1}(y_m)}(y_k)=0$ for all $k\neq m$ and the same is true for $z_k$. Then the measured efficacy parameter reads after integration over all $z_k$ and $y_k$ with $k\neq m$:
\begin{equation}
\begin{aligned}
\gamma_m = \int dz_m\int dy_m p_{\lambda^\dagger(y_m)}(z_m) p_m(y_m|z_m) e^{\beta \left(\delta H_{\lambda^\dagger_{m+1}(y_m)}(y_m)-\delta H_{\lambda^\dagger_{m+1}(y_m)}(z_m)\right)}~.
\end{aligned}
\end{equation} 
The integral of $y_m$ splits into two parts: one in which we alter the potential ($y_m>0$) and one where we do nothing ($y_m<0$):
\begin{equation}
\label{eq appendix splitting of the integral BD eq}
\begin{aligned}
%
&\left<e^{\beta (W_m^\dagger[\yb^\dagger|y_m]-W^\dagger[\zb^\dagger|y_m])}\right>_{\zb^\dagger,\yb^\dagger} = \int dz_m \int\limits_{-\infty}^0 dy_m p_{\lambda^\dagger(y_m)}(z_m) p_m(y_m|z_m)\\
%
&  + \int dz_m \int\limits_{0}^\infty dy_m p_{\lambda^\dagger(y_m)}(z_m) p_m(y_m|z_m) e^{\beta \left(\delta H_{\lambda^\dagger_{m+1}(y_m)}(y_m)-\delta H_{\lambda^\dagger_{m+1}(y_m)}(z_m)\right)}~.
%
\end{aligned}
\end{equation} 
The conditional probability $p_m(y_m|z_m)$ is again assumed to be  Gaussian with a standard deviation of $\sigma_m$ (see Eq.~(\ref{eq gaussian measurement eq})). Moreover, the probability $p_{\lambda^\dagger(y_m<0)}(z_m)$ (no feedback) is the canonical distribution of the harmonic potential centered at $\mu_{\lambda(0)}$ and width $f_{\lambda(0)}$ and the probability $p_{\lambda^\dagger(y_m\geq 0)} (z_m)$ (feedback) is the canonical distribution centered at $\mu_{\lambda(t_f)}$ and width $f_{\lambda(t_f)}$, because we are in equilibrium before applying the backwards protocol. Then the first term of Eq. (\ref{eq appendix splitting of the integral BD eq}) becomes $1/2$ after integration of $z_m$ and $y_m$. If feedback ist applied ($y_m>0$) it holds 
\begin{equation}
\delta H_{\lambda^\dagger_{m+1}(y_m)}(y_m)-\delta H_{\lambda^\dagger_{m+1}}(z_m) = (f_{\lambda(0)}-f_{\lambda(t_f)})(y_m^{ 2}-z_m^{ 2}) - 2f_{\lambda(t_f)}\mu_{\lambda(t_f)}(z_m -y_m)~.
\end{equation}
Then after integration over $z_m$ and $y_m$ of the second part of Eq.~(\ref{eq appendix splitting of the integral BD eq}) one arrives at Eq.~(\ref{eq brownian motion feedback eq}).

\subsection{Derivation for the two level system under feedback}\label{subsec appendix two level feedback}
Here, we derive the analytic expression of the MJE for a driven TLS under feedback, Eq.~(\ref{eq changed Jarzynski feedback two level eq}). We again assume that the driving protocol changes continuously and depends on the measurement outcome $y_m$ at time $t_m$. Then, the measured efficacy parameter of the TLS is given by
\begin{equation}
\label{eq appendix derivation two level feedback continous step in between eq}
\begin{aligned}
%
\gamma_m&= \sum\limits_{\zb^\dagger}\sum\limits_{\yb^\dagger}\mathcal P_{\lambda^\dagger(y_m)}[\zb^\dagger]\prod\limits_k \left[(1-\eta)\delta_{y_k,z_k}+\eta(1-\delta_{y_k,z_k})\right] e^{\beta(\delta\varepsilon_{\lambda_{k+1}(y_m)}(y_k)-\delta\varepsilon_{\lambda_{k+1}(y_m)}(z_k)}~.
\end{aligned}
\end{equation}
Since the driving protocol depends on $y_m$, we can write $\gamma_m$ as:
\begin{equation}
\begin{aligned}
%
&\gamma_m=\sum\limits_{\zb^\dagger} \sum\limits_{y_m} \mathcal P_{\lambda^\dagger(y_m)}[\zb^\dagger] \left[(1-\eta)\delta_{y_m,z_m}+\eta(1-\delta_{y_m,z_m})\right]e^{\beta\left(\delta \varepsilon_{\lambda_{m+1}(y_m)}(z_{m})-\delta \varepsilon_{\lambda_{m+1}(y_m)}(y_{m})\right)} \\
%
& \times\prod_{k\neq m}\left[\sum\limits_{y_k}(1-\eta)\delta_{y_k,z_k}+\eta(1-\delta_{y_k,z_k})\e^{\beta\left(\delta \varepsilon_{\lambda_{k+1}(y_m)}(z_k)-\delta \varepsilon_{\lambda_{k+1}(y_m)}(y_k)\right)}\right]~.
\end{aligned}
\end{equation}
Summing over all $y_k$ results in 
\begin{equation}
\label{eq appendix before assuming eq}
\begin{split}
\gamma_m	=&~	\sum\limits_{\zb^\dagger} \sum\limits_{y_m} \mathcal P_{\lambda^\dagger(y_m)}[\zb^\dagger] \left[(1-\eta)\delta_{y_m,z_m}+\eta(1-\delta_{y_m,z_m})\right]e^{\beta\left(\delta \varepsilon_{\lambda_{m+1}(y_m)}(z_{m})-\delta \varepsilon_{\lambda_{m+1}(y_m)}(y_{m})\right)} \\
		& \times\prod_{k\neq m}\left( (1-2 \eta)+\eta \left[1+e^{\beta (\delta \varepsilon_{\lambda^\dagger_{i+1}}(\bar z_i)-\delta \varepsilon_{\lambda^\dagger_{i+1}}( z_i))}\right]\right)	\\
		\approx&~	\sum\limits_{\zb^\dagger} \sum\limits_{y_m} \mathcal P_{\lambda^\dagger(y_m)}[\zb^\dagger] \left[(1-\eta)\delta_{y_m,z_m}+\eta(1-\delta_{y_m,z_m})\right]	\\
		& \times\prod_{k\neq m}\left( (1-2 \eta)+\eta \left[1+e^{\beta (\delta \varepsilon_{\lambda^\dagger_{i+1}}(\bar z_i)-\delta \varepsilon_{\lambda^\dagger_{i+1}}( z_i))}\right]\right).
\end{split}
\end{equation}
For the last step we approximated  
\begin{equation}
\exp\left\{\beta(\delta\varepsilon_{\lambda_{m+1}(y_m)}(y_m)-\delta\varepsilon_{\lambda_{m+1}(y_m)}(z_m)\right\}\approx 1
\end{equation}
for the single point at $k = m$. This is justified because the final integral does not depend on 
the value of a single point as long as we change the protocol continuously. 
Following the same intermediate steps as 
in~\ref{subsec appendix derivation two level no feedback continuous} we arrive at 
\begin{equation}
\begin{aligned}
\gamma_m &\approx \sum\limits_{z_m,y_m}\left[(1-\eta)\delta_{y_m,z_m}+\eta(1-\delta_{y_m,z_m})\right]\left[p_{\lambda^\dagger(y_m)} (z_m)\right.\\
&\left.\times \exp\left\{-\beta\eta \int dt~\dot\omega^\dagger_{\lambda(y_m,t)}(p_{e,\lambda^\dagger(y_m)}(t)-p_{g,\lambda^\dagger(y_m)}(t))\right\}\right]~.
\end{aligned}
\end{equation}
Finally, by summing over $y_m$ we arrive at Eq.~(\ref{eq changed Jarzynski feedback two level eq}).

For an instantenous change of the driving protocol, where we assume that the Hamiltonian of the TLS is constant before and after the quench at time $t_m$, $\delta \varepsilon_{\lambda_{i+1}(y_m)}(z_m)$ and $\delta \varepsilon_{\lambda_{i+1}(y_m)}(y_m)$ are the only terms different from zero. Then, Eq.~(\ref{eq two level system feedback quench eq}) follows immediately from evaluating the sum over $y_m$ in Eq.~(\ref{eq appendix derivation two level feedback continous step in between eq}).

\end{document}